\documentclass[a4paper,11pt]{article}
\pdfoutput=1 

\usepackage{jheppub} 
\usepackage{amsmath}

 \graphicspath{{figures/}}

\newcommand{\coll}{{{\rm coll} }}

\newcommand{\obs}{{{\rm obs} }}
\newcommand{\inst}{{{\rm inst} }}
\newcommand{\xsep}{x_{\rm sep}}
\newcommand{\Ncoll}{\bar{N}_{\rm coll}}

\definecolor{darkgreen}{RGB}{0,200,0}

\usepackage{soul} 

\title{\bf Simulating the universe(s) II: phenomenology of cosmic bubble collisions in full General Relativity}

\author[a]{Carroll L. Wainwright,}
\author[b,c]{Matthew C. Johnson,}
\author[a]{Anthony Aguirre,}
\author[d]{and Hiranya V. Peiris.}

\affiliation[a]{SCIPP and Department of Physics, University of California \\ Santa Cruz, CA, 95064, USA}
\affiliation[b]{Department of Physics and Astronomy, York University \\ Toronto, On, M3J 1P3, Canada}
\affiliation[c]{Perimeter Institute for Theoretical Physics \\ Waterloo, Ontario N2L 2Y5, Canada}
\affiliation[d]{Department of Physics and Astronomy, University College London \\ London WC1E 6BT, U.K.}

\emailAdd{cwainwri@ucsc.edu}
\emailAdd{mjohnson@perimeterinstitute.ca}
\emailAdd{aguirre@scipp.ucsc.edu}
\emailAdd{h.peiris@ucl.ac.uk}

\abstract{Observing the relics of collisions between bubble universes would provide direct evidence for the existence of an eternally inflating Multiverse; the non-observation of such events can also provide important constraints on inflationary physics. Realizing these prospects requires quantitative predictions for observables from the properties of the possible scalar field Lagrangians underlying eternal inflation. Building on previous work, we establish this connection in detail. We perform a fully relativistic numerical study of the phenomenology of bubble collisions in models with a single scalar field, computing the comoving curvature perturbation produced in a wide variety of models. We also construct a set of analytic predictions, allowing us to identify the phenomenologically relevant properties of the scalar field Lagrangian. The agreement between the analytic predictions and numerics in the relevant regions is excellent, and allows us to generalize our results beyond the models we adopt for the numerical studies. Specifically, the signature is completely determined by the spatial profile of the colliding bubble just before the collision, and the de Sitter invariant distance between the bubble centers. The analytic and numerical results support a power-law fit with an index $1< \kappa \lesssim 2$. For collisions between identical bubbles, we establish a lower-bound on the observed amplitude of collisions that is set by the present energy density in curvature. }

\begin{document} 
\maketitle
\flushbottom

\section{Introduction}

The last decade has seen great progress in bringing the idea of the ``multiverse" from pure theory (at best, and pure speculation at worst) to a testable scientific theory with quantitative predictions for observables. Perhaps the most predictive modern version of the multiverse is ``false vacuum" eternal inflation (see, e.g., Ref.~\cite{Aguirre:2007gy}). In this model, our observable universe is housed inside a bubble, which was itself nucleated from an inflating false vacuum. Because the rate at which bubbles are nucleated is outpaced by the exponential expansion driven by the energy density of the false vacuum, the false vacuum persists indefinitely: inflation is eternal. There are a number of potentially observable traces of this bubbly origin, including: negative curvature (see, e.g., Ref.~\cite{Guth:2012ww} for a recent discussion), modifications of the cosmic microwave background (CMB) temperature and polarization power spectra at the largest scales (see, e.g., Ref.~\cite{Bousso:2013uia} for a recent discussion), and the relics of collisions between bubbles (see, e.g., Ref.~\cite{Aguirre:2009ug,Kleban:2011pg} for a review). Of these effects, bubble collision relics are perhaps the most distinctive observable feature of eternal inflation; this is our focus.

A substantial body of previous work has explored the theory behind bubble collisions~\cite{Hawking:1982ga,Wu:1984eda,Gott:1984ps,Garriga:2006hw,Moss:1994pi,Chang_Kleban_Levi:2009,Chang:2007eq,Czech:2010rg,Freivogel_etal:2009it,Freivogel:2005vv,Gobbetti_Kleban:2012,Kleban_Levi_Sigurdson:2011,Aguirre:2007an,Aguirre:2007wm,Aguirre:2008wy,Johnson:2010bn,Johnson:2011wt,Wainwright:2013lea,Salem:2012,Easther:2009ft,Giblin:2010bd,Hwang:2012pj,Larjo:2009mt}, and observational searches using data from the {\it Wilkinson Microwave Anisotropy Probe} (WMAP) satellite have put constraints on the number and strength of possible features consistent with the observed CMB~\cite{Feeney_etal:2010dd,Feeney_etal:2010jj,Feeney:2012hj,McEwen:2012uk,Osborne:2013hea}. An important missing component of this discussion has been a direct link between the observables and the scalar field Lagrangian underlying a particular model of eternal inflation. To determine what properties of the models can be, and have been, constrained by observations, it is necessary to know quantitatively what theoretical models of eternal inflation predict. 

Recently, in Ref.~\cite{Wainwright:2013lea} we presented the proof-of-principle that this connection can indeed be made by performing fully relativistic simulations of the collision between bubbles containing realistic inflationary cosmologies. The outcome of a bubble collision is entirely determined by the underlying scalar field Lagrangian, and the de Sitter invariant separation between the colliding bubbles (the ``kinematics"). The scalar field Lagrangian dictates the properties of the colliding bubbles and the details of the inflationary cosmology, and the kinematics determine the centre of mass energy of the collision. Causality dictates that after the collision, each bubble interior is split by a ``causal boundary" into regions affected and not affected by the collision. How exactly the affected regions are disturbed is determined by the underlying model and the kinematics. However, when the same scalar field that makes up the bubble also drives inflation inside it, the basic effect of the collision is to advance or retard the evolution of the inflaton in the future of the collision, giving rise to a perturbation in the comoving curvature $\mathcal{R}$. All cosmological observables of interest can be extracted from the comoving curvature perturbation $\mathcal{R}$. Because the collision spacetime enjoys an $SO(2,1)$ symmetry, all the physics can be captured by performing simulations in $1+1$ dimensions.\footnote
{This neglects potential symmetry-breaking fluctuations; see Ref.~\cite{Wainwright:2013lea} for a discussion of caveats.}

This drastic simplification makes it computationally feasible to simulate collisions very accurately for a variety of kinematics in a variety of models. Focusing on regions near the edge of the causal boundary, the spacetime can be approximated as a perturbed Friedmann-Robertson-Walker (FRW) universe in the comoving gauge. 

In Ref.~\cite{Wainwright:2013lea} we presented a method for computing the comoving curvature perturbation $\mathcal{R}$ using a set of geodesics numerically evolved through the simulated collision. For the model chosen, the numerically reconstructed $\mathcal{R}$ in the vicinity of the causal boundary were well-described by a power law in the proper distance from the collision, with a power law index $\kappa$ that decreased with increasing bubble separation (lying between $2 < \kappa < 3.6$). This previous work was a proof of principle for a fiducial model, leaving open questions of how generic our results were for different models of the scalar field potential. 

In this paper, we perform a systematic numerical study of the phenomenology of bubble collisions in models where there is only a single scalar field with a canonical kinetic term. Even with this restriction, there is an infinite-dimensional parameter space to explore, corresponding to the infinite number of single-field potentials. The prediction for observables depends on the post-nucleation slow-roll inflationary period, the post-inflationary cosmology, the properties of the cosmology in the colliding bubble, and the properties of the potential barriers surrounding the false vacuum. To arrive at a parameter space of manageable size, we fix the first three, and focus on variations in the potential barriers. 

Many (arguably most, under some measure) of these potentials would predict either no bubbles at all or a vanishing number of observable collisions, and hence cannot be constrained by searches for bubble collisions. In addition, our assumption of $SO(2,1)$ symmetry will be invalidated for potentials that predict a large number of observable collisions. (See Ref.~\cite{Kozaczuk:2012sx} for a study of the many-bubble regime). We therefore focus on models that predict approximately one observable collision. With these restrictions, we consider two representative models: a predictive model with only one free parameter, and a more generic model with up to five free parameters, which provides enough freedom to model a generic potential barrier. In the generic model, we allow for multiple decay channels from the false vacuum, and therefore collisions between identical and non-identical bubbles. Performing many simulations, we construct a general picture of the phenomenology of $\mathcal{R}$ produced by observable bubble collisions.   

Building on insights gained from the numerical simulations, we build a set of analytic models for the collision spacetime which accurately predict the comoving curvature perturbation produced by collisions. The numerics and the analytic models single out the parameters in the potential that are phenomenologically relevant. Our results therefore lay the foundation for constraining the fundamental Lagrangian underlying eternal inflation directly from cosmological datasets. The ultimate goal of this program is a set of cosmic-variance-limited constraints on the properties of the potential underlying eternal inflation. In future work, we will provide predictions for the signature of bubble collisions for our fiducial models in the CMB (temperature and polarization) and large scale structure (LSS) data, and make comparisons with upcoming datasets. 

\section{From Lagrangians to Perturbations}\label{sec:lagtopert}

\begin{figure}
   \centering
   \includegraphics[scale=0.4]{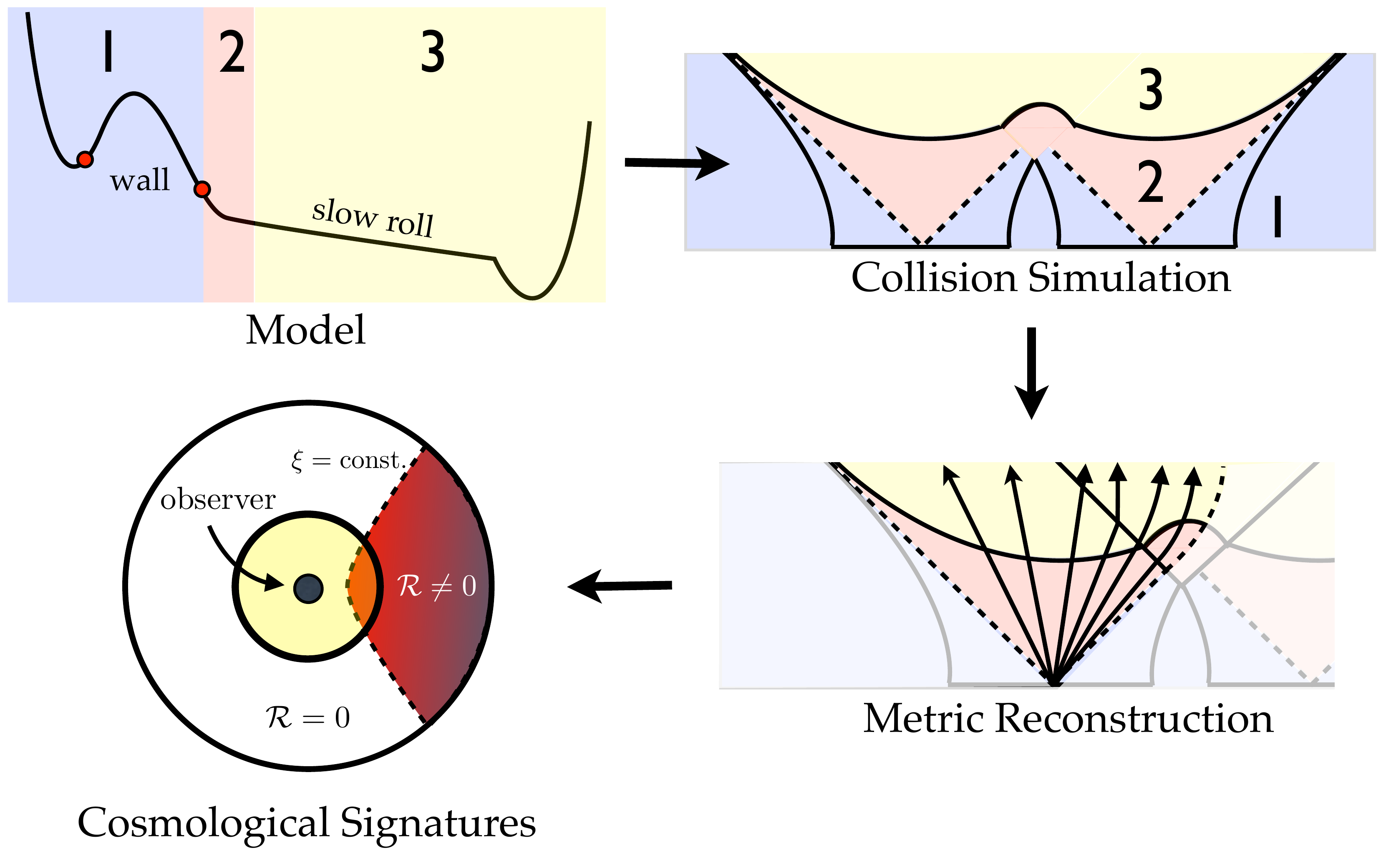} 
   \caption{{ A schematic of the path from a scalar field Lagrangian to the comoving curvature perturbation.} One begins by specifying a scalar-field potential with two or more vacua. With only two vacua (upper left panel), only collisions between identical bubbles are possible. Region 1 on the potential describes the properties of the bubble wall, while regions 2 and 3 describe the cosmological evolution inside of each bubble as depicted in the upper right panel. Simulations are performed to construct the full collision spacetime in a set of global coordinates. Using data from the simulation, we reconstruct the perturbed FRW metric inside the observation bubble by evolving a set of geodesics through the simulation (lower right panel). A gauge transformation then allows us to extract the comoving curvature perturbation $\mathcal{R}$ late in the inflationary epoch. Evolving the comoving curvature perturbation, the observable universe is split into a region that is, and a region that is not, affected by the collision. The lower left panel depicts the surface of last scattering inside the observation bubble in a reference frame where the observer is at the origin of coordinates. The collision boundary follows a line of constant $\xi$ (this coordinate is defined by the metric Eq.~\ref{eq:ximetric} ). This observer's past light cone intersects the collision boundary, and would map on to a disc on their CMB sky.  
   }
   \label{fig:summary_figure}
\end{figure}

We provide a lightning overview of the methods used to find the comoving curvature perturbation $\mathcal{R}$, starting from a specific scalar field Lagrangian. A much more detailed discussion of the methods, including tests of the code we employ, can be found in Ref.~\cite{Wainwright:2013lea}.

The first step in the process is to find the initial conditions for two colliding bubble universes. The bubbles are assumed to be described by Coleman-de Luccia instantons~\cite{Coleman:1977py,Coleman:1980aw}, which can individually be found by using an overshoot-undershoot method (we use the \texttt{CosmoTransitions} package~\cite{Wainwright:2011kj}). To combine initial conditions for two bubbles on the same initial surface, we simply sum their field values, assuming that the overlap is small enough such that they do not affect each other's shape. We then compute metric variables from the constraints. This procedure is valid because the geometry approaches an anisotropic foliation of flat space at early enough times. We designate one bubble as the observation bubble --- it must contain a phenomenologically viable slow-roll inflationary cosmology --- and the other bubble as the collision bubble.

We then simulate the collision using the fully general-relativistic equations of motion on an adaptive grid, which is necessary to accurately resolve the bubble walls and the collision shock fronts. The simulation coordinates are defined in terms of the false vacuum Hubble parameter $H_F$, with an $SO(2,1)$ symmetric metric
\begin{equation}
\label{eq:metric}
H_F^2 ds^2 = -\alpha(N, x)^2 \ dN^2 +  a(N,x)^2 \ \cosh^2 N \ dx^2 + \sinh^2N \ (d\chi^2 + \sinh^2\chi d\varphi^2).
\end{equation}
In the false vacuum state $\alpha=a=1$, reproducing the closed slicing of de Sitter space. The $N$ coordinate measures the approximate number of $e$-foldings in the false vacuum, and the spatial coordinate $x$ is periodic with period $2\pi$. The joint evolution of the metric functions $\alpha(N,x)$ and $a(N,x)$ and the scalar field(s) is determined by the Einstein and scalar field equations, subject to the $SO(2,1)$ symmetry. The configuration is invariant under rotations about $x$ and boosts perpendicular to the $N-x$ plane.

Once a simulation is complete, we reconstruct the metric in a set of cosmological coordinates that we define as a grid of geodesics emanating from the observation bubble origin. The metric is by construction in synchronous gauge. Each geodesic is labeled by its initial trajectory $\xi$: $\frac{dN}{d\tau} = \cosh\xi$ and $\frac{dx}{d\tau} = \sinh\xi$, where $\tau$ is the proper time along the geodesics. If there was no collision (and hence no perturbations), the resulting spacetime would be 
an open Friedmann-Robertson-Walker (FRW) universe, which can be described by the metric
\begin{equation}\label{eq:ximetric}
H_F^2 ds^2 = -d\tau^2 + a(\tau)^2 [d\xi^2 + \cosh^2\xi (d\rho^2 + \sinh^2\rho\ d\varphi^2)],
\end{equation}
where $a(\tau)$ is the time-dependent scale factor. The $(\xi,\rho,\varphi)$ spatial coordinates are a hyperbolic generalization of cylindrical coordinates. In the presence of a collision, we compute the perturbations around the FRW metric by differencing the collision simulation with a simulation of only the observation bubble. Empirically, we find that the trace-free components of the spatial metric perturbations are negligible, in which case we can describe the perturbed spacetime by the synchronous gauge curvature perturbation $D^{\rm syn}$ and the perturbation in the field $\delta \phi$.

Any one observer will have causal access to a finite-sized region on any constant-time hypersurface. Locating the observer at the origin of coordinates, an observer's past light cone at last scattering is a sphere of radius $\Delta\xi_\obs = 2 \sqrt{\Omega_k}$, where $\Omega_k$ is the fraction of the energy density in spatial curvature. Observers who are far inside the unperturbed FRW region see nothing, and observers deep within the collision region presumably see long-wavelength deviations from homogeneity (we will report on the signature seen by such observers in future work). Observers who are within a distance $\Delta\xi_\obs$ of the collision boundary will see their observable universe split into a portion which is affected by the collision and a portion that is not. This gives rise to  distinctive observational signatures, such as azimuthally symmetric features in the CMB~\cite{Aguirre:2007gy}. We concentrate on such observers in this paper.

Since we are primarily interested in the perturbation near the boundary, the perturbed quantities in the synchronous gauge, $D^{\rm syn}$ and $\delta \phi$, are small. This makes the application of linear perturbation theory valid, and we perform a linear gauge transformation to obtain the comoving curvature perturbation
\begin{equation}\label{eq:comovingcurvature}
\mathcal{R} = D^{\rm (syn)} + H \frac{\delta \phi}{\partial_\tau \phi_0},
\end{equation}
where $H$ is the background Hubble parameter and $\phi_0 (\tau)$ is the evolution of the field in the background FRW metric. All cosmological observables of interest can be obtained from the portion of the primordial comoving curvature perturbation within the observable universe. 

\section{Reference Models}

\subsection{Potential definitions}

In order to perform simulations, we must choose a scalar field potential. This maps on to the properties of the false vacuum and bubbles, as well the properties of the slow-roll inflationary cosmology inside  bubbles. 

For the inflationary cosmology, we restrict our study to scalar potentials with an $m^2\phi^2$ slow-roll phase. The (undisturbed) observation bubble interior is homogeneous by symmetry, and begins dominated by negative curvature. A sufficient number of $e$-folds is therefore necessary to dilute the observed curvature $\Omega_k$ down to a phenomenologically acceptable level. We fix the number of $e$-foldings\footnote
{Both the total number of $e$-folds and the details of reheating determine the observable curvature, so there is some uncertainty in our mapping of the choice $\phi_* = 3 M_{\rm Pl}$ to the observed $\Omega_k$.} 
to be $N_* \approx 57$, corresponding to an initial field value $\phi_* \approx 3 M_{\rm Pl}$.
To obtain the correct amplitude for stochastic fluctuations in the curvature perturbation, we set $m \approx 1.3 \times 10^{-6} M_{\rm Pl}$. 

The potential barriers between the false vacuum and true vacua fixes the decay rate(s) of the false vacuum and the properties of the bubble walls. We study two types of barriers: a Gaussian bump barrier, which is very restrictive and has predictive phenomenology, and a quartic potential barrier, which is more general. In all cases, the inflationary minimum is placed at $\phi_0 = 3 M_{\rm Pl}$ away from the false vacuum minimum, which roughly sets the initial field value for the inflationary cosmology. 

Since perturbations from bubble collisions freeze out, changes to the late-time slow-roll portion of the inflationary potential do not change the resulting comoving curvature perturbation $\mathcal{R}$. We can therefore extrapolate our results to varying $\Omega_k$, as long as the potential near $\phi_*$ is held fixed. In addition, the numerical simulations are invariant under an overall rescaling of the potential. Therefore, with $\phi_*$ fixed only, hierarchies between $m$ and the scales characterizing the barrier and false vacuum should affect the final results.

We now describe our two models for the potential in more detail.

\subsubsection{Gaussian bump barrier}\label{sec-gaussbump}

\begin{figure}[t] 
   \centering
   \includegraphics{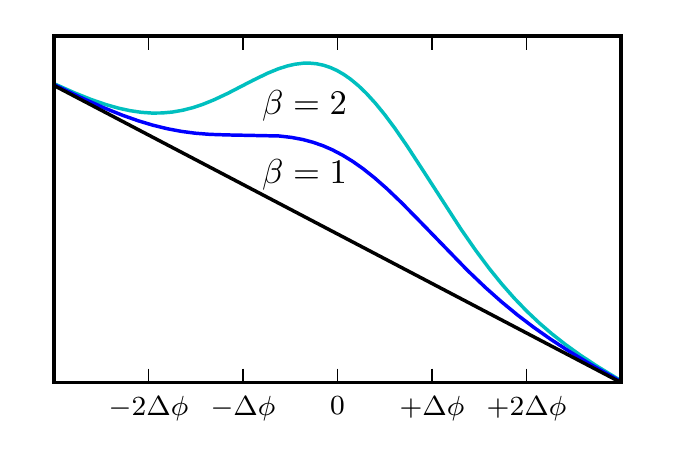} 
   \caption{The Gaussian bump barrier. At $\beta=1$, there is a stationary point but no metastable minimum. Increasing $\beta$ gives rise to a false vacuum minimum. 
   }
   \label{fig:gaussian_bump_potential}
\end{figure}

Our first model, the Gaussian bump, has the potential:
\begin{eqnarray}
V(\phi) = \frac{1}{2}m^2(\phi-\phi_0)^2 + C e^{-\phi^2 / 2 \Delta\phi^2}.
\end{eqnarray}
We set the location of the bump at $\phi=0$ with the inflationary minimum at $\phi_0 = 3 M_{\rm Pl}$. The height of the bump is given by $C$, but it is more convenient to parameterize the height relative to the slope of the quadratic term. Let $C \equiv \beta | m^2\phi_0\, \Delta\phi\, e^{1/2}|$ define a new parameter $\beta$. Then in the limit that $|\phi_0| \gg \Delta\phi$ (such that the slope of the quadratic term 
is approximately constant under the bump), $\beta=1$ corresponds to a bump with a single stationary point but no maximum or minimum (see Fig.~\ref{fig:gaussian_bump_potential}). We therefore restrict to the range $\beta> 1$. Rescaling $\Delta\phi$ while holding $\beta$ fixed is a rescaling of both the bump's height and width, whereas changing $\beta$ changes the qualitative features of the barrier.

\subsubsection{Quartic barrier}\label{sec-quarticbarrier}

The second potential is a quartic polynomial barrier matched to the slow-roll inflationary potential such that the field and it's first derivative are continuous everywhere. 
This has more free parameters than the Gaussian bump model, allowing for an independent tuning of the height, width, and shape of the barrier.

We define the quartic barrier in terms of its first derivative,
\begin{eqnarray}
V'(\phi) = \lambda \phi(\phi-\phi_{\rm top})(\phi-\phi_{\rm bot}),
\end{eqnarray}
so that the potential is given by
\begin{eqnarray}\label{eq:quarticbarrier}
V(\phi) = \lambda\left( \frac{1}{4}\phi^4 - \frac{\phi_{\rm top}+\phi_{\rm bot}}{3}\phi^3 + \frac{\phi_{\rm top} \phi_{\rm bot}}{2}\phi^2\right).
\end{eqnarray}
The metastable minimum is at $\phi = 0$, the top of the potential barrier is at $\phi_{\rm top}$, and the bottom of the barrier is at $ \phi_{\rm bot}$. We define the parameter 
\begin{equation}
\omega \equiv  \phi_{\rm top} /  \phi_{\rm bot},
\end{equation}
which describes the scale-independent barrier shape.
In order for $V( \phi_{\rm bot}) < V(0)$, $\omega$ must be less than $1/2$. Different barriers with identical values of $\omega$ are related to each other through rescaling of the height and width. 

\begin{figure}[t] 
   \centering
   \includegraphics{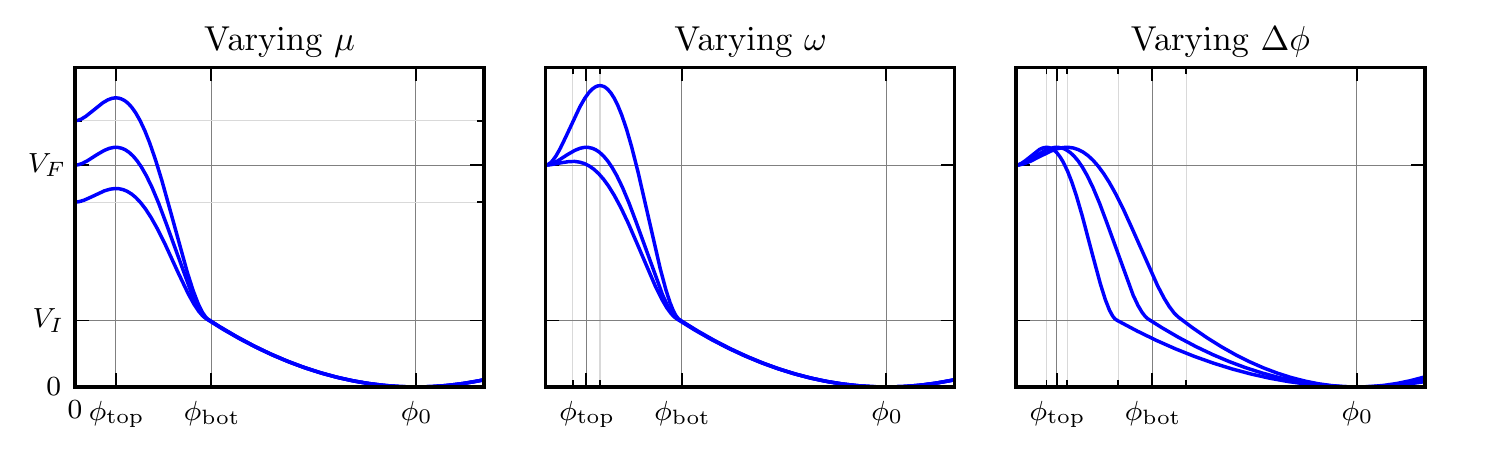} 
   \caption{Examples of a quartic barrier matched on to an $m^2\phi^2$ potential. 
   Each of the three panels shows variations in $\mu$, $\omega$, or $\Delta\phi$ while holding the other two barrier parameters fixed. Any model which produces both Coleman-de Luccia instantons and slow-roll inflation will necessarily have a large hierarchy between $\Delta\phi = \phi_{\rm bot}$ and $\phi_0$, which, for visualization, is not shown here. Although in this exaggerated depiction the value of $m$ varies appreciably, for the cases below, where $\Delta\phi \ll \phi_0$, variations in the barrier parameters will have negligible effects upon the slow-roll quadratic part of the potential.
}
   \label{fig:quartic_barrier}
\end{figure}

The quartic barrier cannot be cleanly added to the slow-roll quadratic potential like the Gaussian bump barrier can. Instead, we define a piecewise-continuous function which matches the barrier and slow-roll potentials together. We do this in the most straightforward way: we match the bottom of the quartic potential barrier\footnote
{The matching does not take place \emph{exactly} at $\phi_{\rm bot}$ --- were that the case, the quartic potential's derivative would be zero and not match the quadratic potential's derivative --- but instead very slightly to the left of $\phi_{\rm bot}$ such that derivatives do match. The exact matching point is found numerically.
} to the top of the slow-roll barrier such that the potential and its first derivative are  continuous everywhere (see Fig.~\ref{fig:quartic_barrier}). This differs from the potentials used in Refs.~\cite{Johnson:2011wt,Wainwright:2013lea}, which allowed for a stationary point at the bottom of the potential barrier.
As stated above, the slow-roll part of the potential is fixed. Therefore, three parameters define the barrier in Eq.~\ref{eq:quarticbarrier}: the barrier width  $\Delta\phi = \phi_{\rm bot}$, the barrier shape parameterized by $\omega$, and the height. We use $\mu \equiv \frac{V_F - V_I}{V_I}$ to parameterize the height, where $V_F = V(0)$ is the energy density in the false vacuum, and $V_I = V(\phi_{\rm bot})$ is the energy density at the top of the slow-roll inflationary region.

Since the quartic potential is defined in a piecewise fashion, it is simple to add a second quartic barrier at negative $\phi$. The definition of the second barrier is the same as the first, but with $\phi_{\rm bot} < \phi_{\rm top} < 0$. The two barriers are made to match at $\phi=0$.

\subsection{Constraining the number of observed collisions}

As discussed in the introduction, analyzing the phenomenology of bubble collisions is only interesting if one can expect there to be at least one contained in the observable universe. Conversely, a precise analysis in which there are many overlapping collisions would require simulations in $3+1$ dimensions. Therefore, we focus on models which have an expectation value of one (or close to one) bubble(s) intersecting the observable portion of the surface of last scattering.  This section demonstrates how this constraint can be enforced, and thus reduces the number of parameters in each potential by one.

The total expected number of observable collisions is given by~\cite{Aguirre:2007wm,Aguirre:2009ug,Freivogel_etal:2009it}
\begin{equation}
\label{eq:Nobs}
\Ncoll \simeq H_F^{-4} \lambda \ \sqrt{\Omega_k} \ \left( \frac{H_I^2}{H_F^2} \right) \approx \frac{1}{H_F^4 r_\inst^4} \ e^{-S} \ \sqrt{\Omega_k} \ \left( \frac{H_F^2}{H_I^2} \right),
\end{equation}
where $H_F$ and $H_I$ are the Hubble parameters in the false vacuum and at the top of the inflationary vacuum, and $\lambda$ is the probability per unit four-volume to nucleate a bubble. This probability is exponentially suppressed by the bounce action $S$, and we have approximated the tunnelling pre-factor as $r_\inst^{-4}$, where $r_\inst$ is the approximate radius of a critical bubble.\footnote{Although the radius of the critical bubble is well defined in the thin-wall limit, we need an operational definition for the smooth profiles we consider. In this paper, we take the critical  radius to be the point at which the field has moved 99\% of the way from the instanton's center to the false vacuum.
}
From constraints on curvature~\cite{Sanchez:2013tga,Ade:2013uln}, we have that $\sqrt{\Omega_k} \lesssim 0.1$.

The bounce action $S$ is given by~\cite{Coleman:1977py}
\begin{equation}
S = 2 \pi^2 \int ds \ r^3 \ \left[ \frac{1}{2} \left( \frac{d\phi}{dr} \right)^2 + V \right],
\end{equation}
where $\phi(r)$ is the solution to the Euclidean equations of motion
\begin{equation}
\frac{d^2\phi}{dr^2} + \frac{3}{r}\frac{d\phi}{dr} = \frac{\partial V}{\partial \phi},
\end{equation}
and interpolates between the false vacuum at $r \rightarrow \infty$ and the instanton endpoint on the true vacuum side of the barrier at $r=0$. We have ignored the effects of gravity in this computation, which is consistent in the limit where the critical radius of the bubble is much smaller than the false vacuum horizon size. By dimensional analysis, the instanton radius scales as $r_\inst \propto \Delta\phi / \sqrt{\Delta V}$, while the action scales as $S \propto \Delta \phi^4 / \Delta V$.

For a given potential barrier with a given barrier height, we can analytically find the barrier width $\Delta\phi_{\Ncoll=1}$ that produces $\Ncoll$ = 1. Rearranging terms in Eq.~\ref{eq:Nobs},
\begin{equation}
(S_0 \Delta\phi^4/\Delta V) e^{ S_0 \Delta\phi^4 /\Delta V} = 
(H_F H_I)^{-2} \sqrt{\Omega_k} \Ncoll^{-1} \Delta V S_0 r_0^{-4},
\end{equation}
where $r_0$ and $S_0$ are the instanton radius and action for a barrier with unit width and height. Then, using the Lambert-W function, we can solve for the barrier width in terms of the other parameters in the potential:
\begin{equation}
\Delta\phi = \left[
W\left( (H_F H_I)^{-2} \sqrt{\Omega_k} \Ncoll^{-1} \Delta V S_0 r_0^{-4} 
\right) \Delta V / S_0 
\right]^{1/4}.
\end{equation}

The calculation for the Gaussian bump barrier differs slightly because the width cannot be varied independently of the height while keeping the barrier shape fixed. Instead, $\Delta V \propto V' \Delta\phi$ and therefore $r_\inst \propto \sqrt{\Delta\phi / |V'|}$ and $S \propto \Delta\phi^3 / |V'|$ for fixed $\beta$, where $V'$ is the slope of the $m^2 (\phi-\phi_0)^2$ potential at $\phi=0$. Performing a similar calculation, we find
\begin{gather}
\frac{3}{2} S_0 (\Delta\phi^3 / V') e^{{(3/2)} S_0 \Delta\phi^3 / V'} = 
\frac{3}{2} \left(\Ncoll^{-1} \sqrt{\Omega_k}\right)^{3/2} (H_F H_I r_0^2)^{-3} \; S_0 V'^2,
\end{gather}
where again $r_0$ and $S_0$ are the instanton radius and action for a Gaussian bump barrier with unit width, fixed $\beta$, and unit $V'$.  This yields
\begin{equation}
\Delta\phi = \left[
W\left(\frac{3}{2} \left(\Ncoll^{-1}\sqrt{\Omega_k}\right)^{3/2} (H_F H_I r_0^2)^{-3} \; S_0 V'^2 \right)
\frac{2 V'}{3S_0}
\right]^{1/3}.
\end{equation}
In the calculations below we use these equations to determine $\Delta \phi$ in terms of the other potential parameters wherever we specify that we have fixed $\Ncoll=1$.

\section{Simulations}
\label{sec:simulations}

\subsection{Code description}

We use the \texttt{CosmoTransitions} package~\cite{Wainwright:2011kj} to calculate the instanton profiles, and an optimized version of the code described in Ref.~\cite{Wainwright:2013lea} to run the collisions. The primary change to the code is in the definition of the monitor function, which determines the spacing of grid points. In Ref.~\cite{Wainwright:2013lea}, the grid density was set to be proportional to $d\phi/dx$, such that the number of grid points in the bubble wall was approximately constant regardless of the wall's width. However, this does not work well with thick-walled bubbles. The problem is that the inside edge of the wall is not a sharp feature --- there is a smooth transition from the center of the bubble to the false vacuum. If we kept the grid density proportional to $d\phi/dx$, there would be a tremendous number of grid points between the center and the nominal wall region, and the simulation would proceed impractically slowly. Instead, we set the grid density to be proportional to the gradient of the canonical momentum conjugate to $\phi$ (see Ref.~\cite{Wainwright:2013lea} for detailed descriptions of simulation variables and the equations of motion), which we have found to be peaked near the outer portion of the bubble wall, where resolution is most important.

In all simulations, we set a minimum grid density of 500 points per unit length (measured in $H_F^{-1}$) along the grid, and set the monitor function such that there are approximately 100 points per bubble wall. After smoothing, a typical grid will have $\sim 3000$ data points during the simulated collision. Each simulation is run until $N=5$, after which point the simulation is truncated to exclude the bubble walls\footnote
{In our previous paper, we were careful to not truncate the simulation until $N=7$. This was important to resolve perturbations very near the bubble wall, which are important for very large kinematic separations between bubbles. Here we only look at separations of $\Delta \xsep = 1$ and 2, so we can truncate the simulation at an earlier time.
} and then continued until $N=50$, which is close to the end of inflation in the center of the observation bubble. All other simulation parameters are the same as in Ref.~\cite{Wainwright:2013lea}.

\subsection{Coordinates and superposition}

As described in Sec.~\ref{sec:lagtopert}, the simulations are performed using  coordinates that cover the entire collision spacetime. The global coordinates are useful for gaining intuition on the outcome of a collision. However, for quantitative comparison with observation, it is necessary to obtain a perturbed FRW metric, which only describes the interior of the observation bubble. We will take advantage of both descriptions of the collision in the discussion below. 

Previous work has established a simple rule for determining the immediate outcome of a bubble collision: just superpose the field profiles of the colliding bubbles as a function of the global coordinates. This ``free-passage" approximation was introduced in Refs.~\cite{Easther:2009ft,Giblin:2010bd} in the absence of gravitational effects (see also Refs.~\cite{Amin:2013dqa,Amin:2013eqa}), and exhibited in simulations incorporating gravity in Refs.~\cite{Johnson:2011wt,Wainwright:2013lea}. This simple rule suggests that in the global coordinates, the collision between identical bubbles leads to an advance of the inflaton, while the collision between non-identical bubbles leads to a delay of the inflaton~\cite{Johnson:2011wt,Gobbetti_Kleban:2012}. Therefore, the profile for single bubbles can be used to gain qualitative and quantitative insight into the effects of a collision. 

\subsection{Gaussian bump model}

\begin{figure}[t] 
   \centering
   \includegraphics{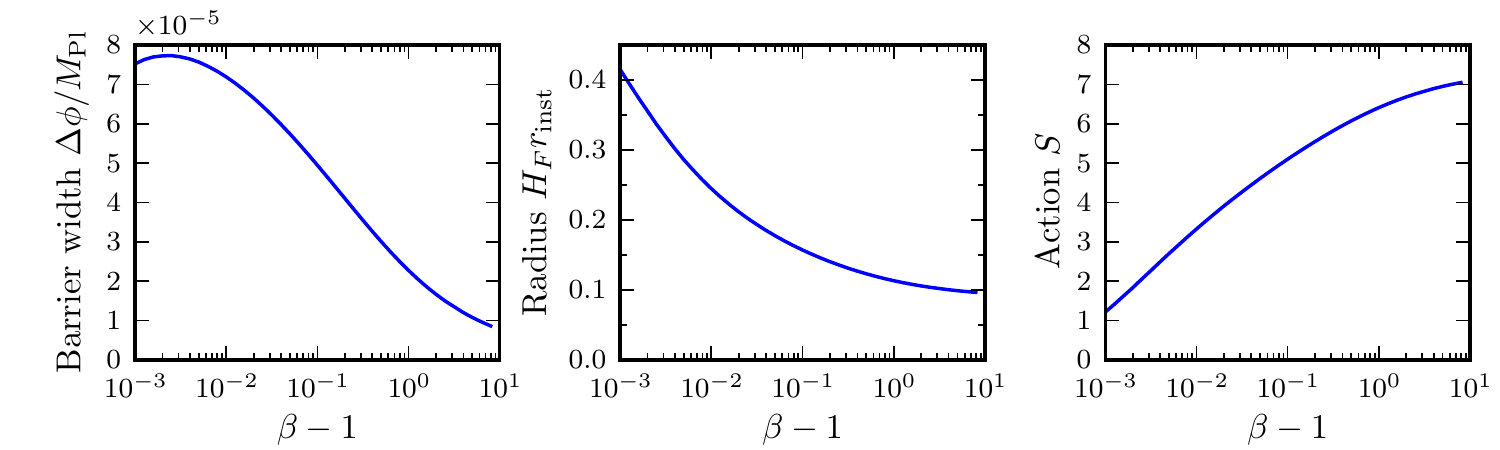} 
   \caption{The width of a Gaussian bump barrier satisfying $\Ncoll=1$ as a function of the barrier height $\beta$, along with the associated instanton's radius and action.
   }
   \label{fig:gaussian_radius_and_action}
\end{figure}

We run an array of simulations for the Gaussian bump model with bump heights ranging from $\beta = 1.001$ to $\beta = 11$.  As $\beta$ gets very close to 1 the bubble radius grows to a substantial portion of the false-vacuum de Sitter radius and the action becomes very small (see Fig.~\ref{fig:gaussian_radius_and_action}). A very large radius precludes placing two well-separated bubbles within the same horizon volume, and a small instanton action undermines the semiclassical approximation we employ for computing the tunneling rate. 
Therefore, we cannot realistically model colliding bubbles with $\beta-1 \ll 1$ and $\Ncoll=1$.
Very large values of $\beta$ are viable, but computationally difficult to simulate: as both the radius and barrier width become very small, a prohibitively large number of grid points are necessary to resolve the bubble wall and the collision region. This restricts us to values $\beta \lesssim 11$.

\begin{figure}[t] 
   \centering
   \includegraphics{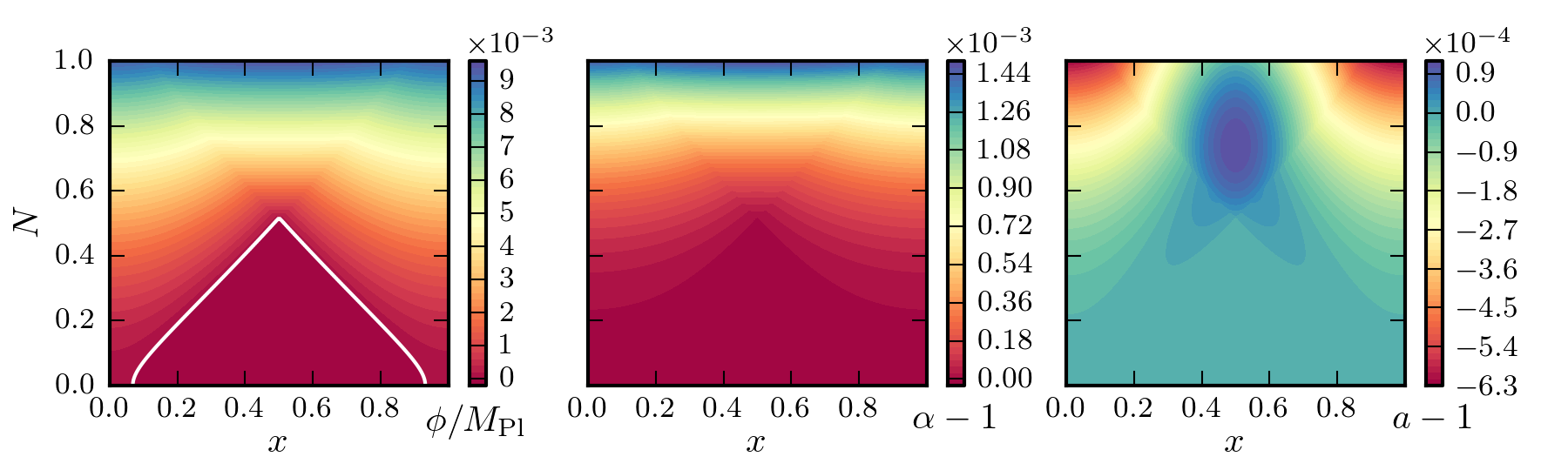} 
   \caption{Contour plots for a simulated bubble collision with a Gaussian bump barrier. The bubbles have a kinematic separation of $\Delta \xsep=1$, and the barrier is defined by $\beta = 1.5$; $\Ncoll=1$. The three plots show the field configuration (left), the metric function $\alpha$ (middle), and the metric function $a$ (right).
   The white line in the left plot shows the exterior boundaries of the two bubbles.
   }
   \label{fig:gaussian_simulation}
\end{figure}

\begin{figure}[t] 
   \centering
   \includegraphics{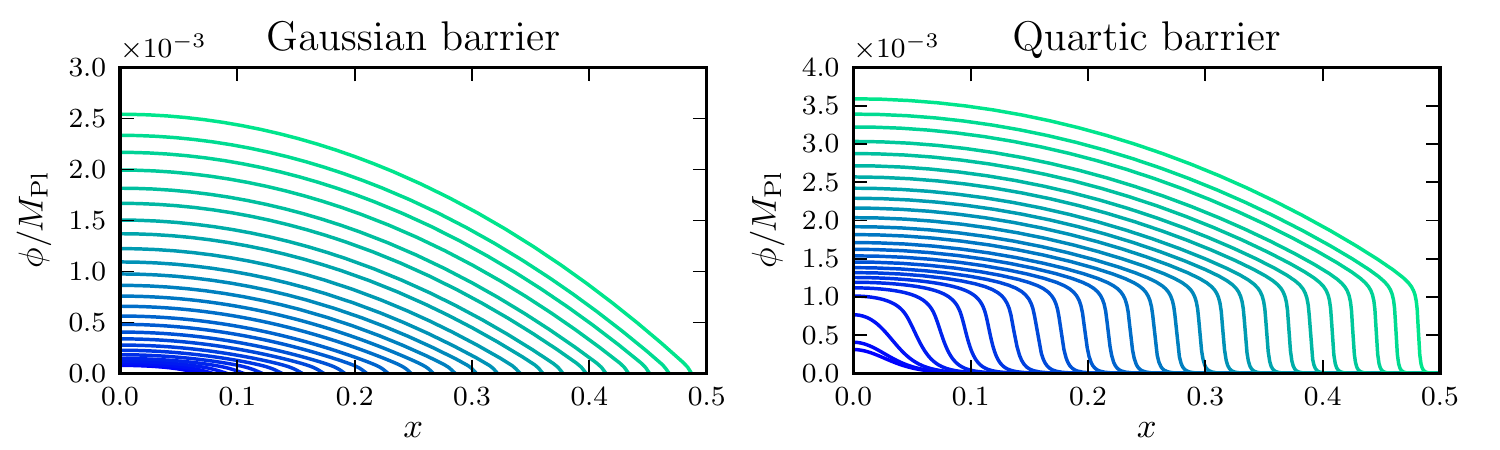} 
   \caption{Early instanton evolution for both the Gaussian bump potential with $\beta=1.5$ and the benchmark quartic barrier potential. Individual lines show field values at constant $N$ time slices, with each time slice linearly spaced between $N=0$ (blue lines) and $N=0.5$ (green lines).
   The Gaussian bump instanton does not exhibit an appreciable barrier wall, while the quartic barrier instanton has a distinct transition from steep wall-like behavior to slow-roll behavior.
   }
   \label{fig:simulation_lines}
\end{figure}

Figure~\ref{fig:gaussian_simulation} shows contour plots of the field and metric functions obtained for a typical simulated collision, with $\beta = 1.5$ and $\Ncoll=1$. The collision boundary appears as a kink in the field profile that propagates into the observation and collision bubbles, and is smoothed by the subsequent evolution. As described above, the properties of individual bubbles can be used to gain some insight into the properties of the collision spacetime. In the left panel of Fig.~\ref{fig:simulation_lines} we show the field profile for a single bubble from the time of nucleation until half a false-vacuum Hubble time. As the bubble grows with time, the field inside the bubble is advancing along the slow-roll potential, giving rise to the growth in amplitude of the bubble profile seen in Fig.~\ref{fig:simulation_lines}. The width of the barrier over the full range in $\beta$ that we study is of order $10^{-5}$ (see Fig.~\ref{fig:gaussian_radius_and_action}). However, the distance that the inflaton rolls in one false-vacuum Hubble time is far larger than this. The profile for essentially all of the Gaussian bump barriers is therefore completely dominated by the slow-roll evolution. 

\begin{figure}[t] 
   \centering
   \includegraphics{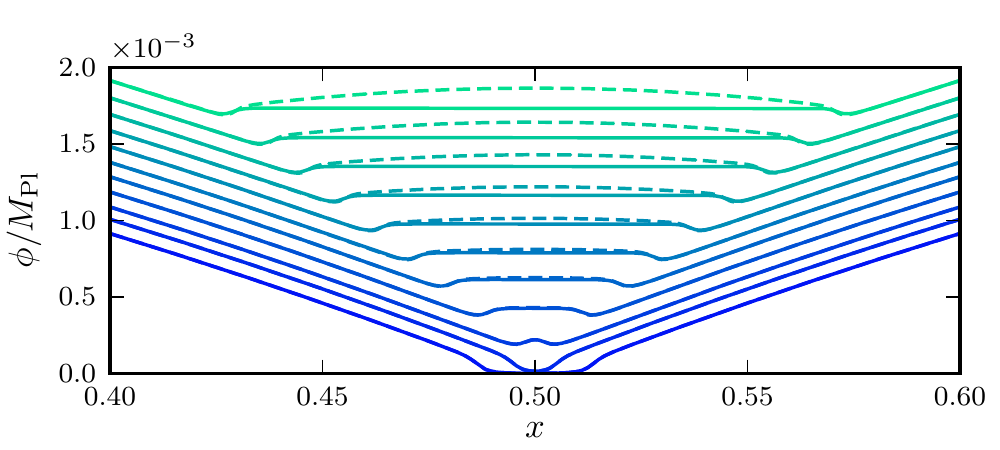} 
   \caption{Several time slices through the simulation of Fig.~\ref{fig:gaussian_simulation} (solid lines), ranging from $N=0.5$ (bottom blue lines) to $N=0.6$ (top green lines), compared to the direct superposition of two copies of the observation bubble (dashed lines). The two bubbles approximately superimpose immediately after the collision. The bubbles each have a roughly linear profile near their outside edges, and the superposition of two opposite linear profiles is a flat region. The flatness persists to later times, even though naive superposition would predict a bump.
   }
   \label{fig:superposition_comparison}
\end{figure}

In Fig.~\ref{fig:superposition_comparison} we show a number of time-slices through the simulation of Fig.~\ref{fig:gaussian_simulation} compared with the direct superposition of two bubble profiles. This differs from most models that have previously been examined, which either assumed vacuum bubbles with no interior cosmology or potentials with stationary points near the instanton endpoint. In both of these cases, the cosmological evolution during a false-vacuum Hubble time is negligible, and the width of the potential barrier was the dominant factor in determining the post-collision field value.

In the cases under study, we clearly see that both the barrier and the post-tunnelling cosmological evolution enter into the free-passage approximation. This suggests that there is a minimum amplitude for the collision between identical bubbles with fixed interior cosmology, since as the barrier width is decreased to zero, the post-collision profile will be determined entirely by the details of the slow-roll evolution. Based on this observation, we predict that the signature of the collisions will not depend very strongly on the value of $\beta$. We comment on the implications of this important observation in greater detail below and in Sec.~\ref{sec-analyt}. 

\begin{figure}[t] 
   \centering
   \includegraphics{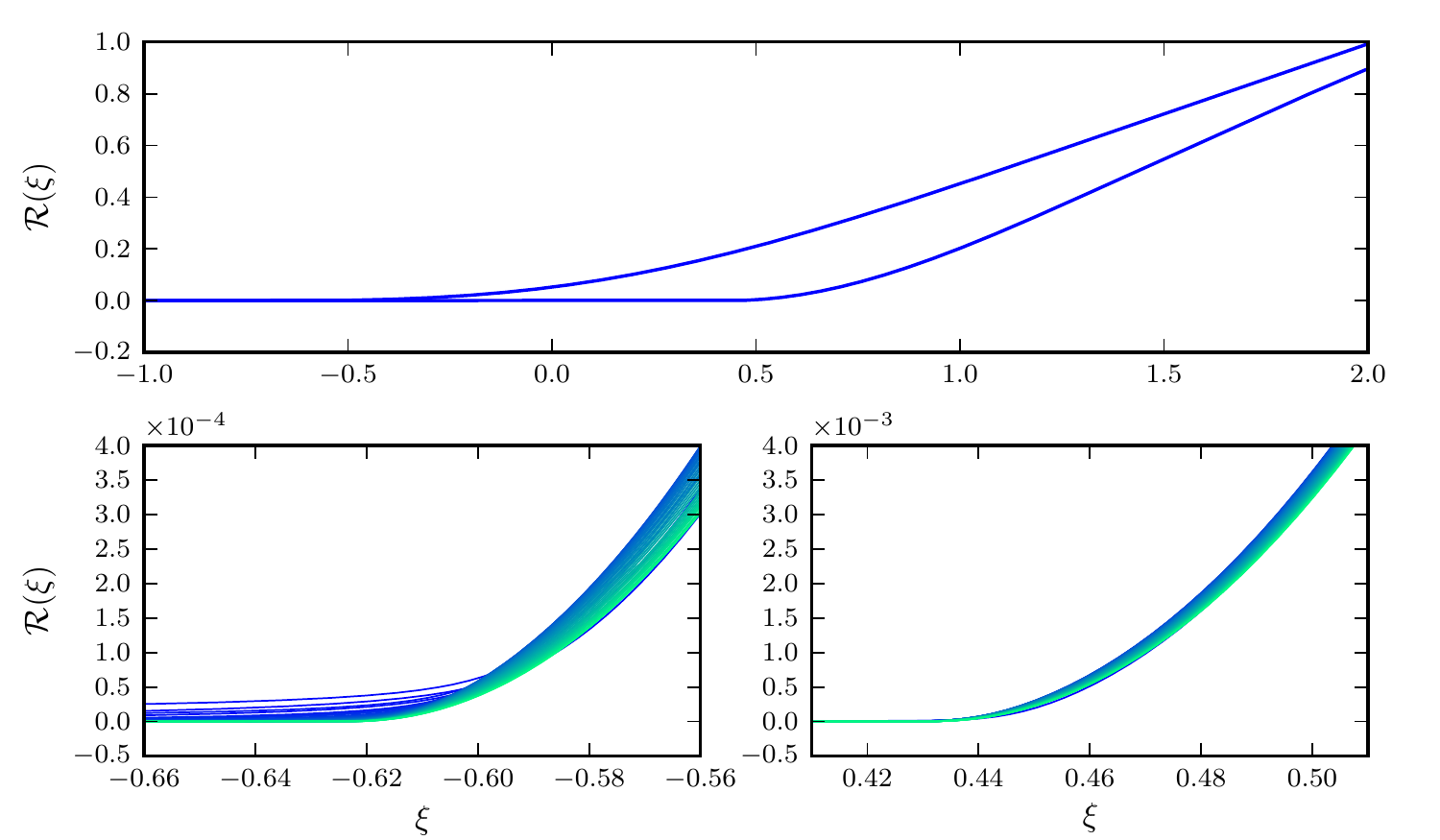} 
   \caption{Comoving curvature perturbations as a function of the hyperbolic coordinate $\xi$ resulting from collisions between bubbles with Gaussian bump potential barriers. The bottom two plots show the same data as the top plot, but zoomed in on the collision regions for $\Delta\xsep =1$ (left) and $\Delta\xsep = 2$ (right). Bluer (greener) lines are for simulations with lower (higher) values of the bump height $\beta$. The perturbation is very weakly sensitive to $\beta$, with differences only apparent in the magnified plots.
   }
   \label{fig:gaussian_perturbations}
\end{figure}

The comoving curvature perturbation as a function of the FRW coordinate $\xi$ for $\Delta x_{\rm sep} = 1, 2$ and $\beta = 1.5$ is shown in the top panel of Fig.~\ref{fig:gaussian_perturbations}. The bottom panels show $\mathcal{R}$ in the vicinity of the collision boundary for all simulated values of $\beta$ for $\Delta x_{\rm sep} = 1$ (left) and $\Delta x_{\rm sep} = 2$ (right). In all cases, there is no sharp collision boundary. Very small values of $\beta-1$ (blue lines) have fairly large initial instanton radii, leading to overlap in the initial conditions for the collision and observation bubbles at $\Delta x_{\rm sep} = 1$, and thus the offset from $\mathcal{R}=0$ seen in Fig.~\ref{fig:gaussian_perturbations}. Fixing  $\Delta x_{\rm sep}$ and varying $\beta$, there is little difference between the perturbations. However, there is a strong dependence on $\Delta x_{\rm sep}$: $\mathcal{R}$ increases by an order of magnitude over a region of fixed size when going from $\Delta x_{\rm sep} =1$ to $\Delta x_{\rm sep} = 2$.

It is helpful to characterize the perturbations in terms of a fitting function. The function needs to be zero outside the collision region and continuous at the collision boundary $\xi_c$. It does not, however, need to be continuously differentiable or analytic at the boundary. In previous work~\cite{Wainwright:2013lea} with a different potential, we found that a power law produced empirically good fits to the collision perturbation near the collision boundary. We use the same fitting function here:
\begin{equation}
\label{eq:fitting}
\mathcal{R}_{\rm fit}(\xi) = A \, \left(\frac{\xi-\xi_c}{\xi_0}\right)^\kappa \Theta(\xi-\xi_c),
\end{equation}
where we take $\xi_0 = 0.05$ as a characteristic length scale. The fitting parameters $A$, $\xi_c$, and $\kappa$ will depend on the model under study as well as the kinematic separation $\Delta\xsep$ between the two colliding bubbles. We calculate the fitting parameters by minimizing 
$\sum_i [\mathcal{R}_{\rm fit}(\xi_i) - \mathcal{R}(\xi_i)]^2$, 
where the sum is over discrete data points linearly spaced in $\xi$. We only include $\xi_i$ that satisfy $2\times 10^{-4} < \mathcal{R}(\xi_i) < 4\times 10^{-3}$. These constraints ensure that we are not attempting to fit numerical noise (which appears at the $10^{-5}$ level), while focusing only on the region near the collision boundary. The quantitative results are only weakly sensitive to the constraints (changing them by a factor of two can result in a 5\% shift in $\kappa$ or a 25\% shift in $A$), and the qualitative results do not change at all.

The comoving curvature perturbation and the fitting parameters are determined by the collision alone, independent of the position of the observer. They are tied to the observer primarily through the observable extent $\Delta\xi_\obs$ of the observer's past lightcone (see Fig.~\ref{fig:summary_figure}, or, for a more detailed explanation of observer coordinate transformations, Fig.~13 in Ref.~\cite{Wainwright:2013lea}). 
For example, if an observer's recombination (or other late-time) surface covers $\Delta\xi_\obs = 0.1$, then the central amplitude of a half-sky bubble would be $\mathcal{R}_{\rm central}  = \mathcal{R}(\xi_c + 0.05) \approx A$.

\begin{figure}[t] 
   \centering
   \includegraphics{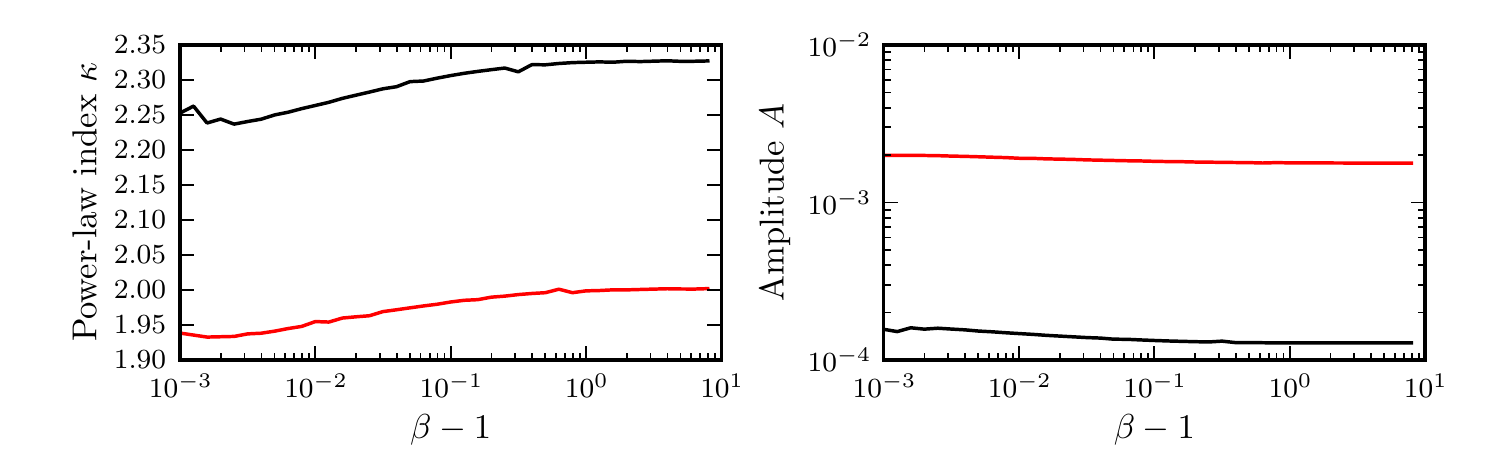} 
   \caption{Best-fit power-law parameters for the comoving curvature perturbations resulting from collisions between bubbles with Gaussian bump potential barriers. Black lines are for bubbles with kinematic separation $\Delta \xsep=1$, and red lines are for $\Delta \xsep=2$.
   }
   \label{fig:gaussian_fits}
\end{figure}

We show the fitting parameters $A$ and $\kappa$ for all values of $\beta$ at representative\footnote{
The probability of nucleating a bubble (per unit space-time volume) with a particular value of $\Delta\xsep$ is $P \propto \sin^3 \Delta\xsep$ (see Ref.~\cite{Wainwright:2013lea}, Eq.~7.24). The values $\Delta\xsep = 1$ and 2 lie on either side of the distribution's peak with relative probabilities of $P/P_{\rm peak} = 0.60$ and 0.75.
}
 $\Delta \xsep = 1, 2$ in Fig.~\ref{fig:gaussian_fits}. 
The power law index decreases slightly with increasing separation, but is nearly flat with $\beta$. All perturbations are closely fit by a power law with index $\kappa \approx 2$. The amplitude increases with increasing separation, and is nearly independent of $\beta$. The Gaussian bump model is therefore highly predictive. There is only one parameter ($\beta-1$) that can be changed while keeping $\Ncoll$=1, and varying this parameter over four orders of magnitude produces less than a 5\% change in the resulting best-fit power-law index and less than a 30\% change in the fitted amplitude. This result agrees with the prediction from superposition of the bubble profiles: since the barrier width is negligible compared to the distance that the field slow-rolls during the first $e$-fold, the perturbation is nearly independent of the parameter controlling the properties of the barrier.

\subsection{Quartic barrier model for identical bubbles}

As discussed above, there are three independent parameters that can vary in the quartic potential barrier: the barrier width $\Delta\phi$, the relative difference in energies between the bottom and top of the barrier $\mu$, and the relative position of the barrier top given by $\omega$. The width can be fixed by the requirement that $\Ncoll=1$. It is convenient to compare different simulations to a common reference point. We pick as a benchmark the potential parameterized by $\mu=0.01$, $\omega=0.054$, and $\Delta\phi = 7.9\times10^{-4} M_{\rm Pl}$ (with these choices, $\Ncoll=1$ for $\sqrt{\Omega_k} = 0.1$).

\begin{figure}[t] 
   \centering
   \includegraphics{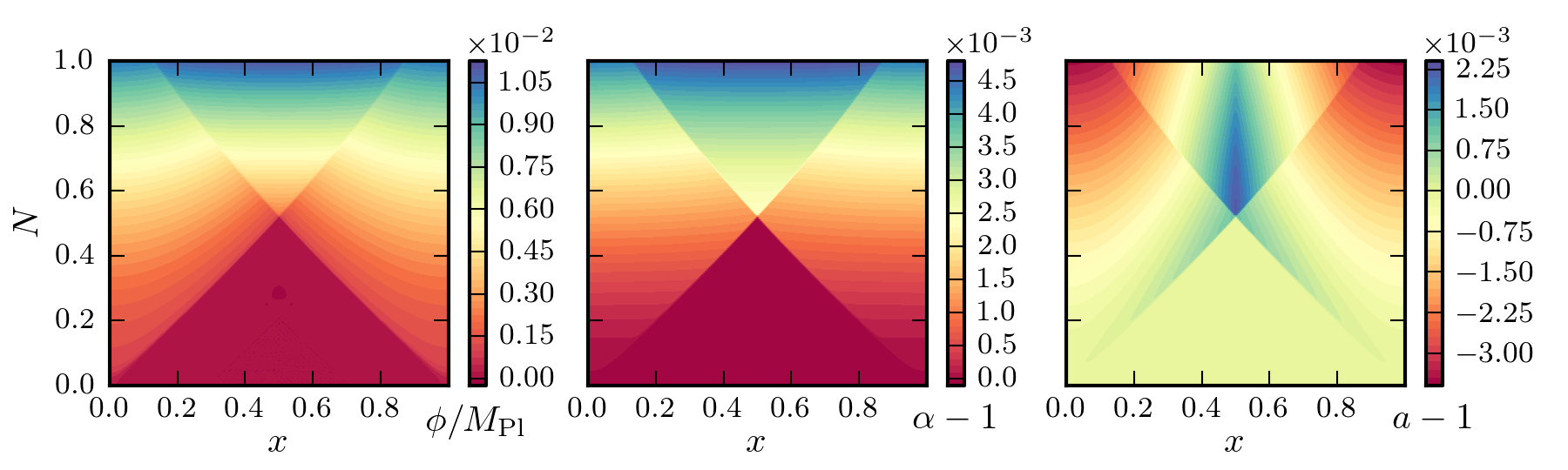} 
   \caption{Contour plots for a simulated bubble collision with the benchmark quartic barrier and $\Delta\xsep=1$. Compared to the Gaussian bump model, the collision boundary is a much sharper feature.
   }
   \label{fig:quartic_BM_simulation1}
\end{figure}

Figure~\ref{fig:quartic_BM_simulation1} shows the simulation output for the benchmark point. Unlike the Gaussian bump simulation, this simulation exhibits far more distinct bubble walls and a sharper shock front emanating from the collision. However, this does not mean that the benchmark bubble is thin-walled; it is not.
The field value at the bubble's center is in the middle of the barrier wall.\footnote
{An instanton's endpoint does not generally sit near the bottom of the potential barrier. 
Smaller values of $\omega$ create smaller potential barriers, leading to more thickly-walled instantons. Conversely, large values of $\omega$ (close to $\omega = \frac{1}{2}$) lead to more thinly-walled instantons.}
When the bubble grows, it first quickly rolls down the remainder of the potential barrier before entering the slow-roll phase of the potential. It is this fast-roll phase that gets length-contracted as the bubble grows, creating a distinct bubble wall. This can be seen in the right panel of Fig.~\ref{fig:simulation_lines}. 
The center of the bubble at $N=0$ is at $\phi = 4.0\times 10^{-4} M_{\rm Pl}$, and the top of the slow roll barrier is at $\phi = 7.9\times 10^{-4} M_{\rm Pl}$, but the field does not slow down until roughly $12.0\times 10^{-4} M_{\rm Pl}$, which also corresponds to the bubble wall's effective field amplitude. 

All three parameters in the potential determine the distance the field rolls before entering slow-roll.
Superposition suggests that the width of the wall, together with the distance in field space between the instanton end-point and where slow-roll begins, most strongly predict the amplitude of the perturbation.
Therefore, all three potential parameters affect the overall amplitude of the perturbation. We can identify several clear trends. The field excursion scales with $\Delta \phi$, so this is certainly an important parameter. Decreasing $\omega$ makes the barrier increasingly asymmetric, driving the instanton endpoint away from a critical point, and increasing the duration of fast-roll. Finally, changing $\mu$ does not change the location of the instanton end-point, and therefore can only affect the total field excursion in fast-roll by increasing the initial velocity of the field as it enters the inflationary portion of the potential. We therefore expect the amplitude to increase only slightly with increasing $\mu$. 

\begin{figure}[t] 
   \centering
   \includegraphics{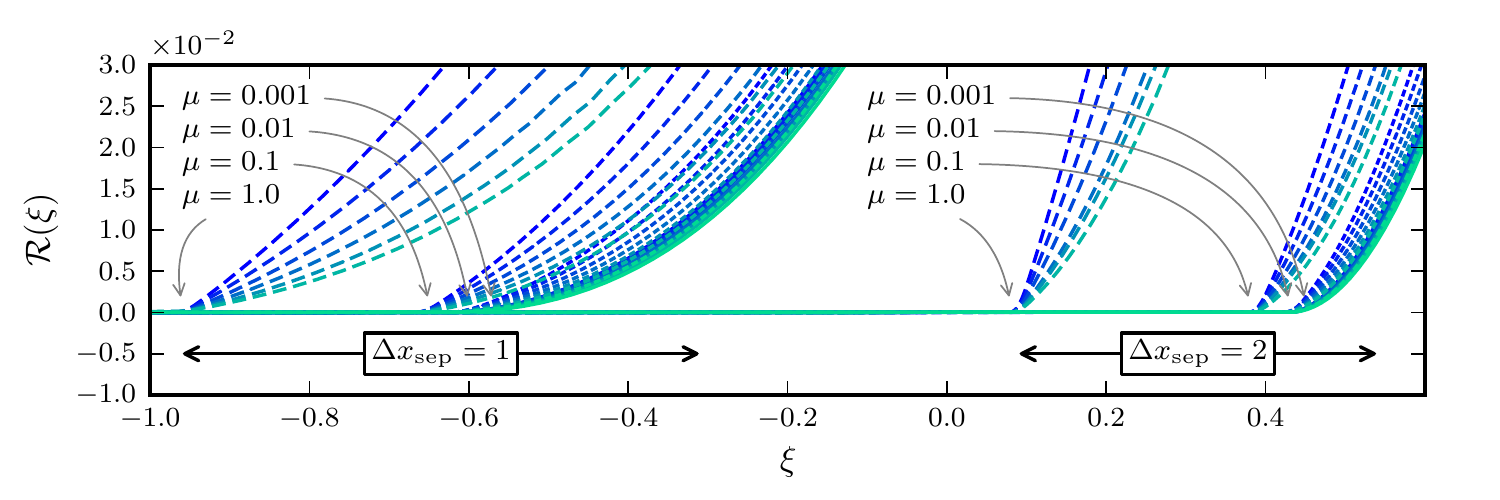} 
   \caption{Comoving curvature perturbations for different collisions with quartic barriers, each satisfying $\Ncoll=1$. The relative barrier height $\mu$ (longer dashes denote larger $\mu$) most strongly affects the location of the collision region, whereas variations in $\omega$ only change the perturbation strength. The value of $\omega$ varies from 0.02 (blue lines) to 0.4 (green lines), with lower values producing larger perturbations.
   }
   \label{fig:quartic_perturbations1}
\end{figure}

\begin{figure}[t] 
   \centering
   \includegraphics{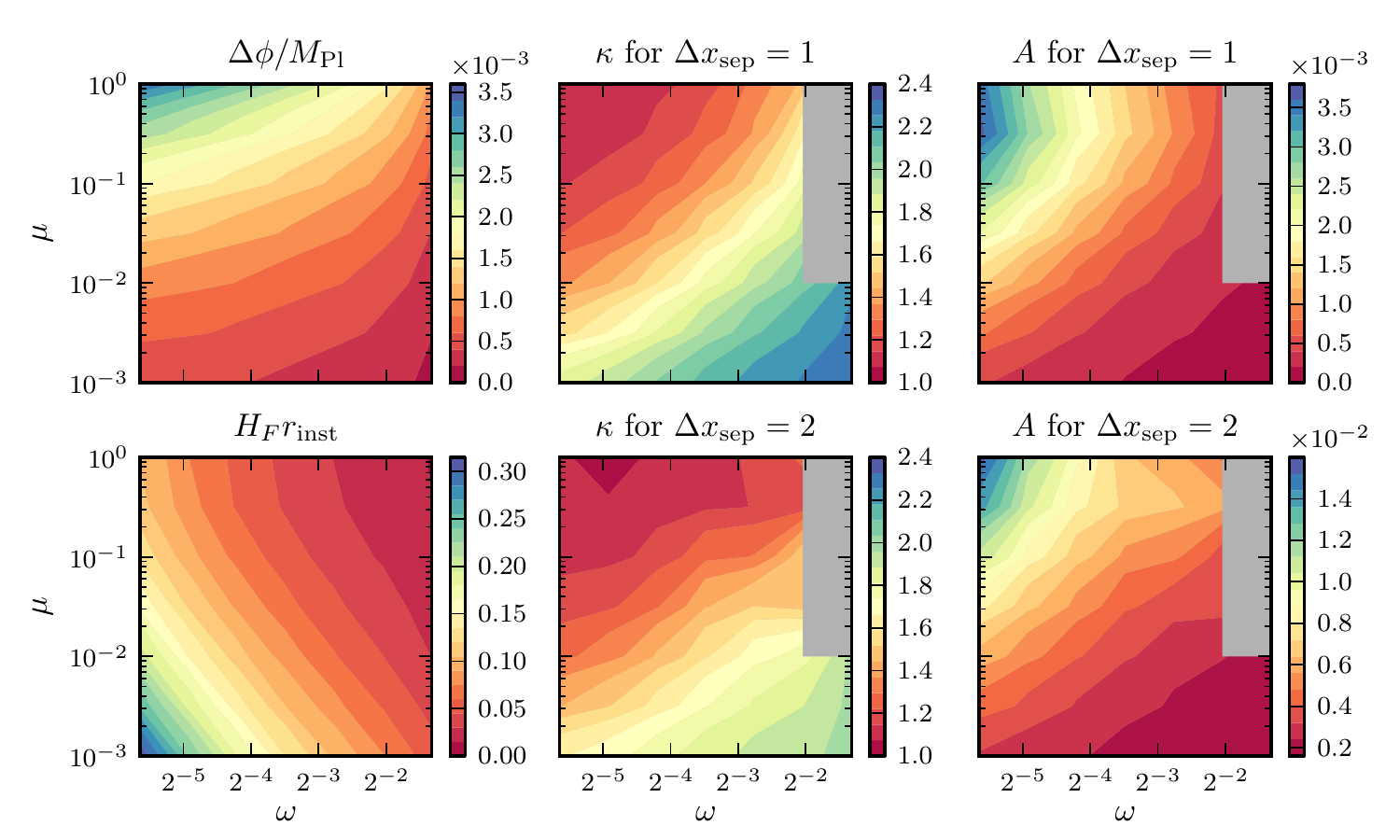} 
   \caption{Barrier width, instanton radius, and fitting parameters $\kappa$ and $A$ for different collisions with quartic barriers. The barrier width $\Delta\phi$ is fixed by $\Ncoll=1$. Gray regions are excluded due to long computation times.
   }
   \label{fig:quartic_fits1}
\end{figure}

To understand quantitatively how the perturbations change with different potential parameters, we simulated many collisions with potential parameters along the $\Ncoll=1$ hypersurface for $\Delta \xsep = 1,2$. The extracted comoving curvature perturbations are shown in Fig.~\ref{fig:quartic_perturbations1}. In Fig.~\ref{fig:quartic_fits1} we show the result of fits in the parameterization of Eq.~\ref{eq:fitting}, along with the barrier width $\Delta\phi_{\Ncoll=1}$ and instanton radius $H_F r_{\rm inst}$ fixed by $\Ncoll = 1$, leaving free the parameters of $\mu$ and $\omega$. 
Note that simulations with very small initial instanton radii are very computationally expensive. A smaller radius means a narrower wall with more length contraction, and narrow walls are more difficult to accurately resolve than thick walls. We therefore do not include results for both large $\mu$ and large $\omega$, and have grayed out these values in Fig.~\ref{fig:quartic_fits1}. It would also be difficult to simulate potentials with large $\mu$ (larger than we show in this paper) for similar reasons. A large $\mu$ corresponds to a large pressure difference across the bubble wall, which leads to more acceleration and generally more length contraction, again making the wall difficult to resolve.

The most obvious effect upon the perturbations comes from changing $\mu$, which shifts the location of the collision boundary. In the thin-wall limit, the collision boundary is given by
\begin{equation}
\label{eq:xi_c}
\xi_c = \log\left[\frac{H_I}{H_F} \tan\left(\frac{\Delta\xsep}{2}\right)\right]
=\log\left[\sqrt{\frac{1}{1+\mu}} \tan\left(\frac{\Delta\xsep}{2}\right)\right],
\end{equation} 
which qualitatively matches our results --- an increase in $\mu$ shifts the boundary to the left. While keeping $\Ncoll$ fixed, increasing $\mu$ also somewhat increases the amplitude and sharpness of the collision (a sharper collision has a lower power-law index $\kappa$), while increasing $\omega$ appears to have the opposite effect. 

However, changing $\mu$  and $\omega$ with $\Ncoll=1$ also indirectly changes $\Delta\phi$, which we expect to have a large effect upon the perturbation amplitude. Superposition suggests that larger $\Delta \phi$ in the collision bubble should correspond to a larger perturbation across the shock front in the observation bubble. Figure~\ref{fig:quartic_fits1} shows that the barrier width is in fact highly correlated with the fitted amplitude. It is anti-correlated with the power-law index $\kappa$, indicating that larger perturbations tend to also have sharper collision boundaries. Interestingly, the instanton radius has very little correlation with either fitting parameter. This is a convenient physical result: computational feasibility restricts the parameter scan to regions with $H_F r_{\rm inst} \sim 0.1$, but since there is no correlation between the instanton radius and the comoving perturbation then this restriction does not bias our results towards a particular observational signature.

\begin{figure}[t] 
   \centering
   \includegraphics{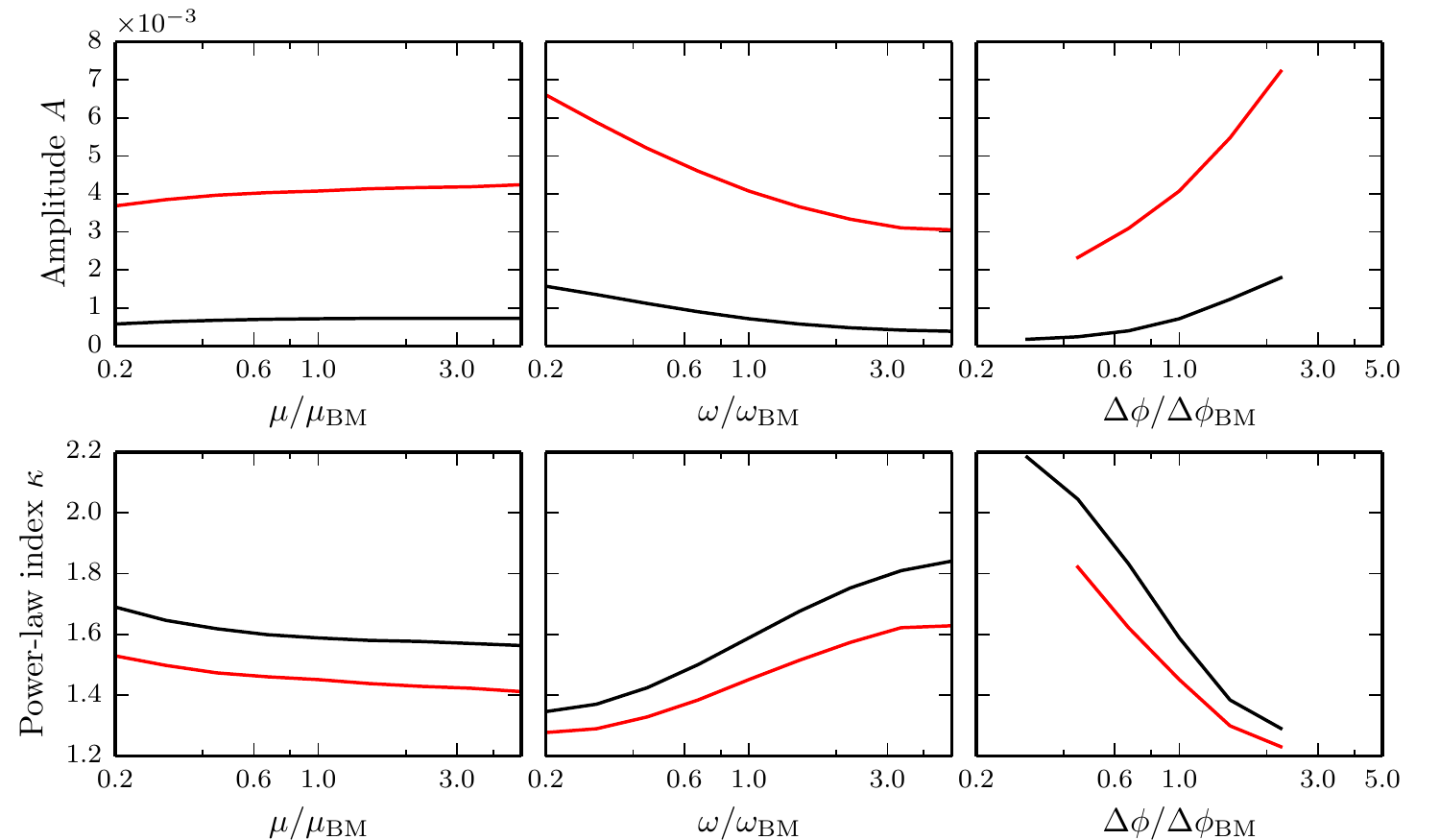} 
   \caption{Perturbation fit parameters for variations in the quartic barrier near the benchmark point, with $\Delta \xsep=1$ (black lines) and $\Delta\xsep=2$ (red lines).
   Simulations with small $\Delta\phi$ are quite computationally intensive, especially for larger $\Delta\xsep$, and are excluded from this plot. Simulations with large $\Delta\phi$ are excluded due to excessive overlap between the initial conditions for the separate instantons.
   }
   \label{fig:quartic_fits2}
\end{figure}

To disentangle the effects of changing $\Delta\phi$ and of changing $\mu$ and $\omega$, we perform one-dimensional scans varying the three parameters individually without the $\Ncoll=1$ requirement. Each scan starts at the benchmark point, with parameters varied by factors of 5 in both directions. The variations in $\Delta\phi$ do not quite cover the full range --- computational constraints limit the range at the low end, while the instanton radius limits the range at the high end (the instantons get too large for large $\Delta\phi$).
Figure~\ref{fig:quartic_fits2} shows the resulting power-law fit parameters. As expected from superposition, variations in $\Delta \phi$ have the largest effect, with steeper and larger amplitude perturbations resulting from wider potential barriers. 
Rescaling the barrier height with $\mu$ has relatively little effect upon the perturbations, although all else being equal, larger differences between the metastable and inflationary phases do yield somewhat stronger perturbations. Making the barrier increasingly asymmetric by decreasing $\omega$ yields steeper and larger amplitude collisions. This is again consistent with the larger superposition of bubble profiles due to the larger excursion in field space between the instanton end-point and the onset of slow-roll for asymmetric potential barriers. 

\subsection{Quartic barrier model for non-identical bubbles}

Finally, we examine the collision phenomenology for non-identical bubbles. We define the observation bubble to be at positive field values with an internal slow-roll cosmology, just as in the previous cases, and add a separate barrier and minimum at negative field values for the collision bubble. We use the quartic barrier and do not include a slow-roll region at negative $\phi$, so the bottom of the collision bubble's potential barrier is a local minimum.

There are two major differences when the collision bubble is at negative field values compared to when both collision and observation bubbles are identical. 
First, the field superposition during the collision is destructive --- the collision retards the field evolution rather than advancing it further down the slow-roll potential. Second, there is a potential barrier between the interiors of the two different bubbles that results in a persistent domain wall. If the collision bubble retards the field all the way back to the original false vacuum (at $\phi=0$), then a pocket of false vacuum forms immediately after the collision. The pocket has its own domain walls which expand, but ultimately collapse due to the pressure gradient. These mini-domain walls  then collide, creating a secondary false-vacuum pocket which expands and recollapses. This  goes on until there is no longer sufficient energy available to return to the false vacuum. This behavior was studied in the very first simulations of bubble collisions~\cite{Hawking:1982ga}.

\begin{figure}[t] 
   \centering
   \includegraphics{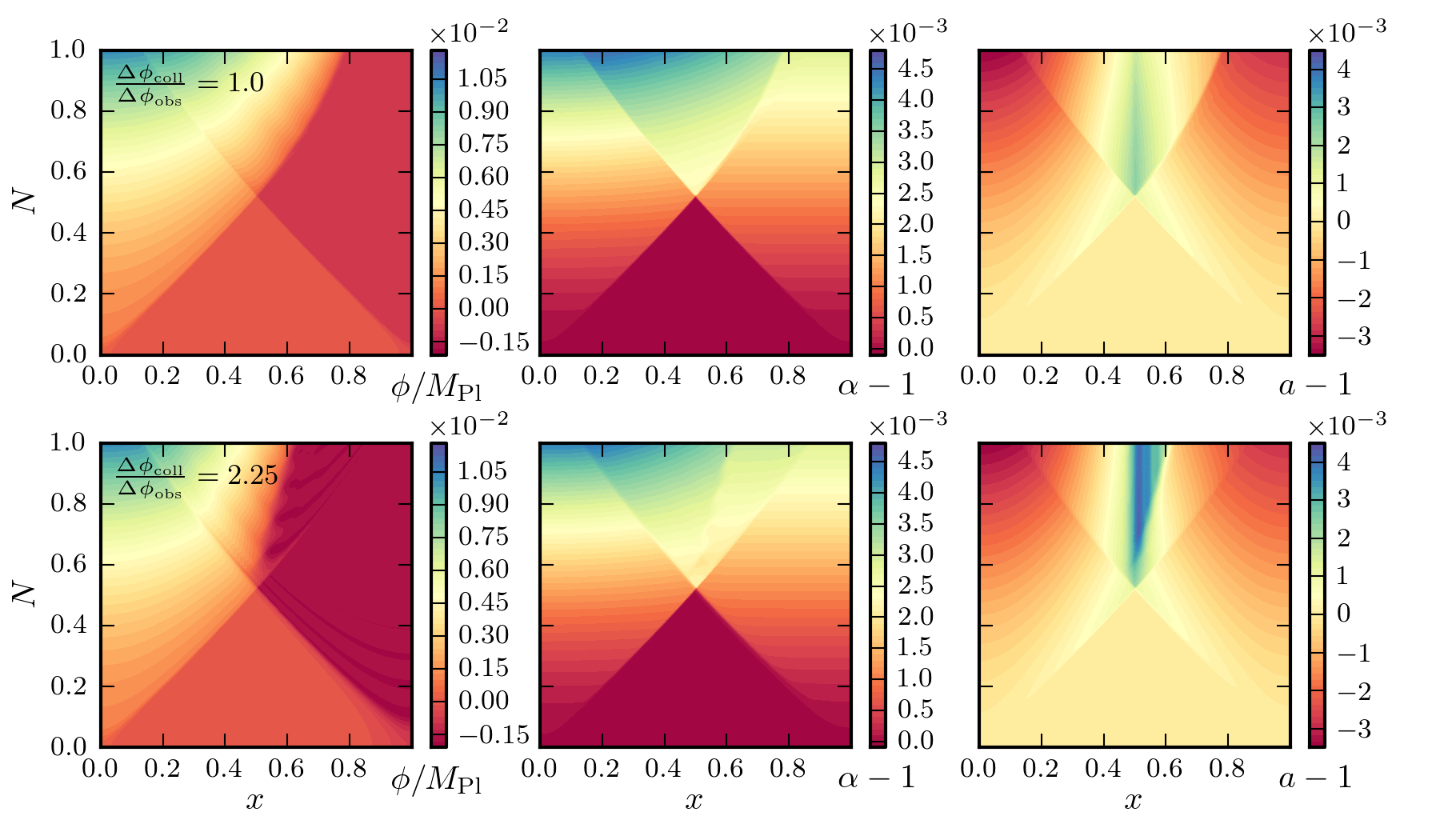} 
   \caption{Simulation contours for collisions between non-identical bubbles with $\Delta\xsep=1$. In the first row the barrier for the collision bubble is the same as the barrier for the observation bubble, but at negative field values. In the second row the collision barrier width is 2.25 times greater than the observation bubble. In both cases the collision bubble lacks a slow-roll interior.
   }
   \label{fig:quartic_nonident_simulation1}
\end{figure}

Figure~\ref{fig:quartic_nonident_simulation1} shows the outputs for two simulations between non-identical bubbles. In both cases the observation bubble (at $x=0$) has the same quartic barrier as that used in the benchmark point in the previous section. In the upper series of plots the collision bubble potential barrier is the mirror image of the observation barrier, whereas in the bottom row the collision barrier has a larger $\Delta\phi$. In both simulations, the field is retarded in the collision region, and both exhibit oscillations in the domain wall. However, the oscillations are much larger in the bottom row where $\Delta\phi_\coll / \Delta\phi_\obs = 2.25$. The larger value of $\Delta\phi_\coll$ causes a deeper cut into the observation bubble, and, since, the bubbles are thick-walled, spreads the domain wall and its oscillatory behavior over a wider region (both physically and in field space). 

\begin{figure}[t] 
   \centering
   \includegraphics{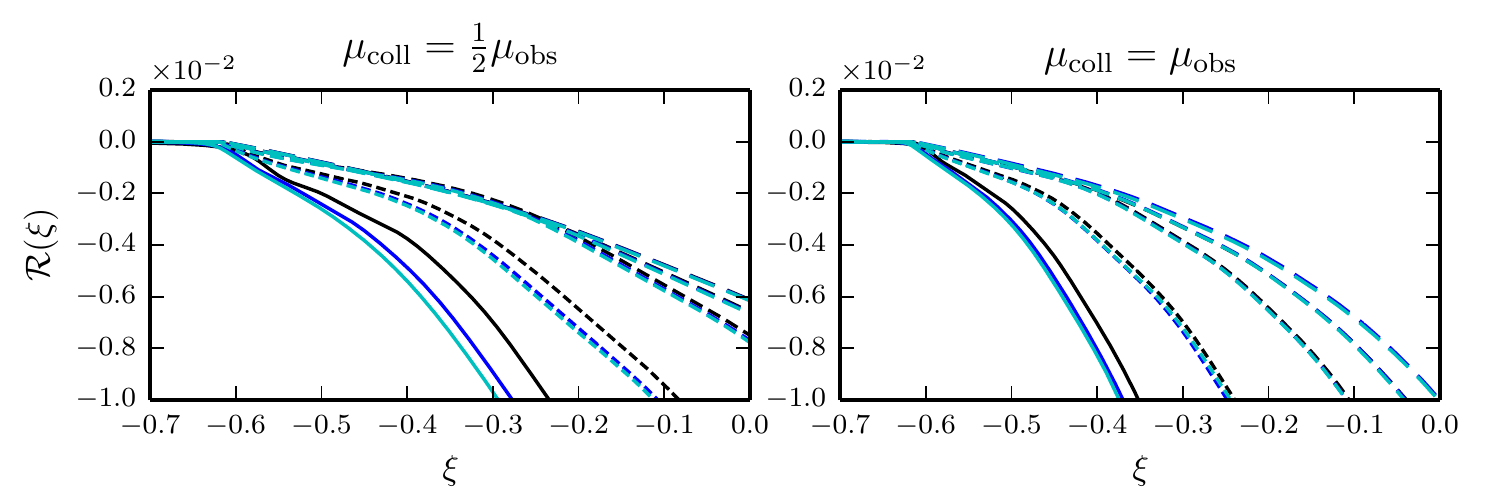} 
   \caption{Perturbations from collisions between non-identical bubbles with $\Delta\xsep=1$. Black, dark blue, and light blue lines indicate $\omega_{\rm coll} / \omega_{\rm obs} = (\frac{3}{2})^{-2}$, 1, and $(\frac{3}{2})^2$ respectively. Different dash lengths indicate different values of $\Delta\phi_{\rm coll} / \Delta\phi_{\rm obs}$, ranging from $(\frac{3}{2})^{-2}$ (longest dashes) to $(\frac{3}{2})^{2}$ (solid lines).
   }
   \label{fig:quartic_nonident_perturbations1}
\end{figure}

The comoving curvature perturbation resulting from the simulations is shown in Fig.~\ref{fig:quartic_nonident_perturbations1}. We simulate an array of non-identical bubble collisions, varying  all three collision bubble quartic barrier parameters while keeping the observation bubble parameters fixed at the benchmark point. We do not enforce $\Ncoll=1$, which will only be true when the two barriers are mirror images of each other. The most distinctive feature of the perturbations is that they are all negative --- a direct result of the retardation rather than advancement of the field inside the observation bubble. The absolute magnitude of the perturbations is comparable to those in the identical bubble case, with the perturbation amplitude resulting from the mirror-image benchmark collision a factor of roughly three lower than that for the identical bubble benchmark collision. The domain wall oscillations appear directly as oscillations in the perturbations. They are most apparent at large $\Delta\phi_\coll$. However, the oscillation wavelength tends to be quite large (with a wavelength in $\xi$ of $\sim 0.1$), and would likely not be apparent at sub-horizon scales. Because of the internal dynamics of the post-collision bubble wall, the power law fit in Eq.~\ref{eq:fitting} is no longer a good description of the perturbation. We therefore do not report best fit values for $A$ and $\kappa$.

Unlike the identical bubble cases, the variations in the relative barrier position $\omega$ of the collision barrier make little difference to the resulting perturbation. Since there is no slow-roll phase inside the collision bubble, any momentum gained by rolling down the potential barrier does not advance the field and change the effective $\Delta\phi$ across the collision bubble wall. The momentum turns into oscillations about the quartic minimum, as seen in the lower left plot of Fig.~\ref{fig:quartic_nonident_simulation1}. On the other hand, changes to $\mu_\coll$ do create large differences in the perturbations. Larger $\mu_\coll$ corresponds to higher pressure inside the collision bubble, which will change the trajectory of the post-collision domain wall, pushing it farther into the observation bubble and resulting in higher amplitude perturbations. Variations in $\Delta\phi_\coll$ have a similar effect as in the identical bubble case: larger $\Delta\phi_\coll$ causes larger field displacements (although now in the negative direction), leading to larger perturbations.

This concludes our discussion of the numerical simulations. As we have seen, the superposition of bubble profiles in the immediate aftermath of the collision provides a good qualitative prediction for the outcome of the collision in the full numerical simulations and the behaviour of $\mathcal{R}$. We now discuss an analytic approximation for $\mathcal{R}$, which provides a quantitative estimate for the comoving curvature perturbation.

\section{Analytic Interpretation of Simulation Results}
\label{sec-analyt}

The full 1+1 dimensional equations of motion are nonlinear and involve both the field and metric functions. However, under the following conditions we can ignore the metric evolution and treat the field equations as linear:
\begin{itemize}
\item The field excursion during the collision and subsequent perturbation freeze-out are small compared to the Planck scale.
\item The potential energy is approximately equal to the de Sitter false-vacuum energy during this same period. This condition, along with the previous one, guarantees that the metric functions are $\alpha\approx a\approx 1$.
\item The slow-roll parameters are approximately constant, so that the potential can be written as $V(\phi) = V(\phi_0) + V'(\phi_0)(\phi-\phi_0)$. (A quartic term in the potential would not ruin linearity in the equations of motion, but it would make finding a solution more difficult.)
\end{itemize}
With these assumptions, the equation governing the field $\phi$ is 
\begin{equation}
\label{eq:linear_eom}
\ddot{\phi} = - \left( \tanh N + \frac{2}{\tanh N}\right) \dot{\phi} + \frac{\phi''}{\cosh^2 N} - \left.\frac{\partial V}{\partial\phi}\right|_{\phi_0}.
\end{equation}
An analytic treatment using a similar set of approximations was performed in Ref.~\cite{Gobbetti_Kleban:2012}. In addition to confirming the main results of this work, we employ a different method of finding solutions and make the following important contributions. 
First, we find exact analytic solutions to the Fourier space version of Eq.~\ref{eq:linear_eom}. 
These mode functions are used to find the evolution of an arbitrary initial field profile (Eq.~\ref{eq:fourier_transform0}). Relying on the fact that the immediate aftermath of a collision can be accurately described by the superposition of bubble profiles, we find analytic expressions for the field perturbation arising both from a steep wall (Eq.~\ref{eq:step_solution}) and from slow-roll inside the observation bubble (Eq.~\ref{eq:phi_lin_solution}). In both cases we derive the dependence on the initial bubble separation. Finally, we derive approximate analytic expressions for the comoving curvature perturbation arising from bubble collisions (Eqs.~\ref{eq:R_analytic_step1} and~\ref{eq:R_analytic_lin1}). These final expressions accurately reproduce the results of our simulations, and can be used to generalize our results to arbitrary single-field Lagrangians. 

\subsection{General solution}

Let us first focus on the homogenous equation ($\partial V/\partial\phi = 0$). Our strategy will be to break an initial condition into its distinct Fourier modes, find the late-time behavior for each mode, and then sum the modes together to find the late-time behavior for an arbitrary initial condition. The homogeneous equation describes the evolution of perturbations on top of a fixed background field configuration. For a given mode $k$, the homogenous part of Eq.~\ref{eq:linear_eom} is
\begin{equation}
\label{eq:homo_eom}
\ddot{\phi} + \left( \tanh N + \frac{2}{\tanh N}\right) \dot{\phi} + \frac{k^2}{\cosh^2 N}\phi = 0.
\end{equation}

This can be rewritten as a Gaussian hypergeometric equation by using the substitution $u = \tanh^2 N$.
The solutions can be found in terms of ordinary hypergeometric functions, which can then be simplified to trigonometric functions. The general solution to Eq.~\ref{eq:homo_eom} is
\begin{equation}
\label{eq:homo_sol}
\phi(N) = C\left[ \cos\left(k\sin^{-1}(\tanh N) - \theta\right) - \frac{k}{\sinh N} \sin\left(k \sin^{-1}(\tanh N) - \theta\right)\right]
\end{equation} 
for arbitrary constants $C$ and $\theta$.

It is helpful to separate the solutions into pieces which either have their zeroth or first derivative vanish at time $N=N_0$. That is, we wish to find solutions $\phi_A$ and $\phi_B$ such that $\phi_A(N_0) = \dot{\phi}_B(N_0) = 1$ and $\dot\phi_A(N_0) = \phi_B(N_0) = 0$. These solutions are given by the constants
\begin{align}
\theta_A(N_0, k) &= k \sin^{-1}(\tanh N_0) - \tan^{-1} [k \sinh (N_0)] \\
\theta_B(N_0, k) &= k \sin^{-1}(\tanh N_0) - \tan^{-1} [ \sinh (N_0)/k] \\
C_A(N_0, k) &= \frac{\sqrt{1+k^2 \sinh^2 (N_0)}}{1-k^2} \\
C_B(N_0, k) &= \frac{\sqrt{1+\sinh^2 (N_0)/k^2}}{1-k^2} \cosh N_0 \sinh N_0.
\end{align}
Both solutions are finite in the limit that $k\rightarrow 1$. At late times, the two solutions reduce to
\begin{align}
\label{eq:phiA_asym}
\phi_A(N\!=\!\infty, N_0, k) &= \frac{1}{1-k^2} \left[ \cos (k\eta) - k\sinh (N_0) \sin (k\eta) \right] \\
\label{eq:phiB_asym}
\phi_B(N\!=\!\infty, N_0, k) &= \frac{\cosh N_0 \sinh N_0}{1-k^2} \left[ \cos (k\eta) - \frac{\sinh (N_0)}{k} \sin (k\eta) \right],
\end{align}
where $\eta \equiv \frac{\pi}{2} - \sin^{-1}(\tanh N_0)$ is the late-time radius of a future light cone starting at $N_0$ as measured with the $x$ coordinate. 

The time dependence of any Fourier mode can  be written as a combination of $\phi_A$ and $\phi_B$:
\begin{equation}
\phi(N,x,k) = \cos\left(k (x-x_0)\right) \phi_A(N, N_0, k) + c_0 k \sin\left(k (x-x_0)\right)\phi_B(N, N_0, k).
\end{equation}
To first order in $N-N_0$, this is equal to $\cos(k(x-x_0) - c_0 k(N-N_0))$, so it describes the complete evolution of a cosine wave with initial velocity $c_0$. To find the evolution of more complicated initial conditions, we must integrate over Fourier modes (see Appendix~\ref{app:fourier_transforms}):
\begin{align}
\label{eq:fourier_transform0}
	\phi(N,x) &= \frac{1}{\pi} \int_0^\infty dk \int_{-\infty}^\infty \phi(N, \tilde{x}) \cos(k(x-\tilde{x})) \, d\tilde{x} \\
		&= \frac{1}{\pi} \int_0^\infty dk \int_{-\infty}^\infty \phi(N_0, \tilde{x}) \left[ \cos (k(x-\tilde{x})) \phi_A + c_0 k \sin(k(x-\tilde{x})) \phi_B \right] \, d\tilde{x},
\label{eq:fourier_transform}
\end{align}
where we assume that each mode has the same initial velocity.

\subsection{Evolving perturbations from bubble collisions}

There are two limiting cases of physical interest for bubble collisions. The first case is for a colliding bubble which has a well-defined wall profile. 
We will now treat the solution to Eq.~\ref{eq:homo_eom} as a perturbation $\delta\phi$ to be added to an unperturbed bubble which satisfies the heterogenous equations of motion.
If we approximate the colliding bubble's profile as a step function, then in the immediate aftermath of the collision where superposition holds, we can set $\delta\phi(N_0, x)$ as a step function moving in the negative $x$-direction at the speed of light:
\begin{align}
\delta\phi_{\rm step}(N_0, x) = \delta\phi_0 \Theta(x).
\end{align}
For now we leave the step function at the origin, but it can easily be translated in the $x$ direction. At the point of collision,\footnote{
From the metric Eq.~\ref{eq:metric}, null geodesics emanating from the origin obey $\tanh N = \pm\sin x$, and, by symmetry, null geodesics from two different bubbles will cross at $x=\Delta\xsep/2$.} $N_0 = \tanh^{-1}[\sin (\Delta\xsep/2)]$, so 
\begin{equation}
\label{eq:eta_xsep}
\eta = \frac{\pi-\Delta\xsep}{2}.
\end{equation}
After integrating Eq.~\ref{eq:fourier_transform} (see Appendix~\ref{app:step_transform}), we find
\begin{eqnarray}
\label{eq:step_solution}
\delta \phi_{\rm step}(N\!=\!\infty, x) = \delta\phi_0 \left\{ \begin{aligned}
&0 &\text{ for }& x \leq -\eta \\
&\frac{1}{2} + \frac{\sin x}{2\sin \eta} + \frac{1}{2}\cot\eta\left(\frac{\cos x}{\sin\eta} - \cot\eta \right)
&\text{ for }& |x| < \eta \\
&1 &\text{ for }& x \geq \eta.
\end{aligned} \right.
\end{eqnarray} 
Note that the step function's dispersion is limited to its future light cone ($|x| < \eta$), as we should expect.
Near the collision boundary ($0 < x+\eta \ll 2\eta$), 
\begin{equation}
\label{eq:step_taylor_series}
\delta \phi_{\rm step}(N\!=\!\infty, x) = \delta\phi_0 \left[\cot\eta \, (x+\eta) 
+ \mathcal{O}(x+\eta)^2\right].
\end{equation}

The second case of interest is when the profile is dominated by a linear contribution,
\begin{align}
\delta\phi_{\rm lin}(N_0, x) = \delta\phi_0' \;x\; \Theta(x).
\end{align}
This type of behavior comes from the slow roll evolution of the bubble interior when there is a negligible potential barrier, as seen in the left plot of Fig.~\ref{fig:simulation_lines}.
Instead of integrating the Fourier transformation equation, it is easier to integrate Eq.~\ref{eq:step_solution} directly. Any arbitrary continuous profile can be expressed as an integral of step functions,
\begin{equation}
\phi(x) = \phi(-\infty) + \int_{-\infty}^\infty \Theta(x-\tilde{x}) \ \phi(\tilde{x})' \ d\tilde{x},
\end{equation}
so the late-time behavior of any perturbation can be given by 
\begin{equation}
\label{eq:any_delta_phi_step}
\delta\phi(N\!=\!\infty, x) =  \int_{-\infty}^\infty \delta\phi_{\rm step}(N\!=\!\infty, x-\tilde{x}) \ \delta\phi(\tilde{x})' \ d\tilde{x},
\end{equation}
where we take $\delta\phi_0=1$ in Eq.~\ref{eq:step_solution} and assume that the perturbation vanishes at $x=-\infty$. Specializing this to the linear profile, we find
\begin{equation}
\label{eq:phi_lin_solution}
\delta\phi_{\rm lin}(N\!=\!\infty, x) = \delta\phi_0' \left\{ \begin{aligned}
&0 &\text{ for }& x \leq -\eta \\
&\frac{x+\eta}{2}(1-\cot^2\eta) + \frac{\sin 2\eta + \sin(x-\eta)}{2\sin^2\eta}
&\text{ for }& |x| < \eta \\
&x + \cot \eta - \eta \cot^2 \eta
&\text{ for }& x \geq \eta.
\end{aligned} \right.
\end{equation}
Near the collision boundary, 
\begin{equation}
\label{eq:lin_taylor_series}
\delta \phi_{\rm lin}(N\!=\!\infty, x) = \delta\phi_0' \left[\frac{1}{2}\cot\eta \, (x+\eta)^2 
+ \mathcal{O}(x+\eta)^3\right],
\end{equation}
which is just the integral of Eq.~\ref{eq:step_taylor_series}.

We can determine the value of $\delta\phi_0'$ analytically for the case of a negligible potential barrier. To do this, we need to solve the heterogenous equation of motion for the unperturbed bubble.
First, one can show that time-like geodesics radiating from the origin of de Sitter space obey $\sinh^2 \tau = \sinh^2 N - \sin^2 x \cosh^2 N$. Surfaces of constant field in unperturbed bubbles correspond to surfaces of constant proper time $\tau$, so
$
\phi'' = \frac{\partial^2\phi}{\partial \tau^2}\left(\frac{\partial\tau}{\partial x}\right)^2 + \frac{\partial\phi}{\partial\tau}\frac{\partial^2\tau}{\partial x^2}.
$
At $x=0$, this reduces to $\phi'' = -\frac{\dot{\phi}}{\tanh N}$, and Eq.~\ref{eq:linear_eom} reduces to 
\begin{equation}
\label{eq:proper_time_eom}
\ddot{\phi} + 3\coth \tau \dot{\phi} + \frac{\partial V}{\partial \phi} = 0,
\end{equation}
where a dot now denotes a proper time derivative. There is nothing special about the geodesic along $x=0$ --- it can be transformed to any other geodesic by a boost --- so Eq.~\ref{eq:proper_time_eom} is valid everywhere inside the bubble.
It can be integrated to 
\begin{equation}
\label{eq:phi_dot_slowroll}
\dot{\phi}(\tau) = - \frac{1}{3}\frac{\partial V}{\partial \phi}\sinh \tau \frac{2+\cosh \tau}{(1+\cosh \tau)^2},
\end{equation}
where again we assume that $\partial V/\partial \phi$ is approximately constant.
Let $\partial_\tau \phi_{\rm sr}$ denote the field's late-time slow-roll speed. 
Then 
\begin{equation}
\label{eq:phi_0_slow_roll}
\partial_\tau \phi_{\rm sr} = -\frac{1}{3}\frac{\partial V}{\partial \phi},
\end{equation}
whereas at small proper time $\partial_\tau \phi(\tau\!\approx\!0) \approx \tau \partial_\tau \phi_{\rm sr}$.
At the bubble's edge, $\frac{\partial\tau}{\partial x} \approx -\frac{1}{\tau}\tan x$. Therefore, the slope of the collision bubble profile at the point of collision is
\begin{equation}
\label{eq:collision_slope}
\partial_x \delta\phi_0 \approx \left[ \frac{\partial\phi}{\partial\tau} \frac{\partial \tau}{\partial x} \right]_{\tau=0} = (\partial_\tau \phi_{\rm sr}) \, \tan \left(\frac{\Delta\xsep}{2}\right).
\end{equation}

Equations~\ref{eq:step_solution},~\ref{eq:phi_lin_solution}, and~\ref{eq:collision_slope} together give an analytic approximation to the late time field perturbation $\delta\phi$ for a colliding bubble with a step-like bubble wall of height $\delta\phi_0$ and an internal cosmology 
which asymptotically approaches a slow-roll value of $\partial_\tau \phi_{\rm sr}$.
These calculations assume that the metric is just that of de Sitter space, and that $\partial V/\partial \phi$ is constant. Both assumptions break down near the end of inflation, but they should hold in the region during which perturbations freeze out, as long as there is neither a large hierarchy between the false-vacuum and inflationary scales nor significant running in the slow-roll parameters prior to freeze-out. 
The total comoving perturbation comprises both a metric term and a $\delta\phi$ term, only the sum of which remains constant. At late times, the $\delta\phi$ term may evolve, but the corresponding gauge-invariant observables will freeze in (where linear perturbation theory holds).

\subsection{Approximating the comoving curvature perturbation} 

Given the approximate late-time field perturbation $\delta\phi$, one can easily calculate the comoving curvature perturbation $\mathcal{R}(\xi)$. In our approximation, there is no metric perturbation inside the observation bubble ($\alpha \approx a \approx 1)$ and the space is approximately de Sitter. Therefore, the comoving curvature perturbation is entirely due to the transformation from synchronous to comoving gauge: $\mathcal{R}(\xi) = H \delta \phi / \partial_\tau \phi_{\rm sr}$. 
For de Sitter space, $H=1$ in our units.
Equation~\ref{eq:any_delta_phi_step} gives the general solution for $\delta\phi$ for any initial bubble profile, while Eq.~\ref{eq:phi_0_slow_roll} determines $\partial_\tau \phi_{\rm sr}$. The last thing needed to find $\mathcal{R}(\xi)$ is the transformation from the global simulation coordinate $x$ to the observer coordinate $\xi$. At asymptotically late times in de Sitter space, the transformation is 
\begin{equation}
\tan x = \sinh\xi.
\end{equation}

We can now construct a relatively simple analytic approximation to the comoving curvature perturbation near the collision boundary.
Let $x_c = \Delta\xsep - \frac{\pi}{2}$ denote the late-time boundary of the collision bubble. Then, after translating the initial perturbation to lie directly between the two bubbles, we can replace
\begin{equation}
x+\eta \;\longrightarrow\; x-x_c 
\approx \frac{dx}{d\xi} (\xi-\xi_c) 
= \cos x_c \, (\xi-\xi_c) 
= \sin(\Delta\xsep) \, (\xi-\xi_c).
\end{equation}
in Eqs.~\ref{eq:step_taylor_series} and~\ref{eq:lin_taylor_series}. From Eq.~\ref{eq:eta_xsep}, we also have $\cot\eta = \tan(\frac{\Delta\xsep}{2})$.
Putting everything together, the approximate comoving curvature perturbations for step-like and linear bubble profiles are given by
\begin{align}
\label{eq:R_analytic_step1}
\mathcal{R}_{\rm step}(\xi) &= \frac{H_I \delta\phi_0}{\partial_\tau \phi_{\rm sr, obs}}
	(1 - \cos \Delta\xsep) (\xi - \xi_c) + \mathcal{O}(\xi-\xi_c)^2 \\
\label{eq:R_analytic_lin1}
\mathcal{R}_{\rm lin}(\xi) &= \frac{\partial_\tau \phi_{\rm sr, coll}}{2\partial_\tau \phi_{\rm sr, obs}}
	(1 - \cos \Delta\xsep)^2 (\xi - \xi_c)^2 + \mathcal{O}(\xi-\xi_c)^3, 
\end{align}
where the amplitude of a step-like profile is given by the difference between the field values at the start of slow roll and at the false vacuum,
\begin{equation}
\delta\phi_0 = \phi_{\rm slow-roll} - \phi_F.
\end{equation}
This is true for both identical and non-identical bubble collisions. However, for non-identical bubble collisions the post-collision domain wall introduces large non-linearities into  the equations of motion. In this case, the analytic approximations are most accurate when $|\delta\phi_{0,\coll}| \ll |\delta\phi_{0,\obs}|$, in which case pockets of false vacuum and other internal degrees of freedom of the post-collision domain wall are not excited. Note that $\delta\phi_0$ has opposite sign for identical and non-identical bubbles. For identical bubble collisions, $\mathcal{R}_{\rm lin}$ is completely independent of the underlying potential: it depends only upon the observer's position $\xi$ and the kinematic separation $\Delta\xsep$. 

A general bubble profile will have both step-like and linear features (with the former due to the finite potential barrier, and the latter due to an internal slow-roll cosmology), 
and the resulting total perturbation will approximately be given by the sum of $\mathcal{R}_{\rm step}$ and $\mathcal{R}_{\rm lin}$. Therefore, depending on the relative contribution from the wall and slow-roll, the perturbation is linear or quadratic in the distance from the collision boundary. Most previous work has assumed a linear dependence. As first rigorously shown in Ref.~\cite{Gobbetti_Kleban:2012}, we have seen that the linear dependence arises from a step-like collision bubble profile. The contribution from slow-roll inside the collision bubble (or an intrinsically large thickness of the collision bubble wall as discussed in Ref.~\cite{Gobbetti_Kleban:2012}), which was not previously considered, is somewhat softer. These analytic results for the $\xi$-dependence of the perturbation are roughly in agreement with the numerical results presented below, where the fitted power law indices varied between $\kappa \simeq 1$ and $\kappa \simeq 2$ in the limit of large $\Delta x$ where the approximations we employ are most accurate. This also agrees with the power law indices obtained in the large $\Delta x$ limit of the model studied in Ref.~\cite{Wainwright:2013lea}.

\begin{figure}[t] 
   \centering
   \includegraphics{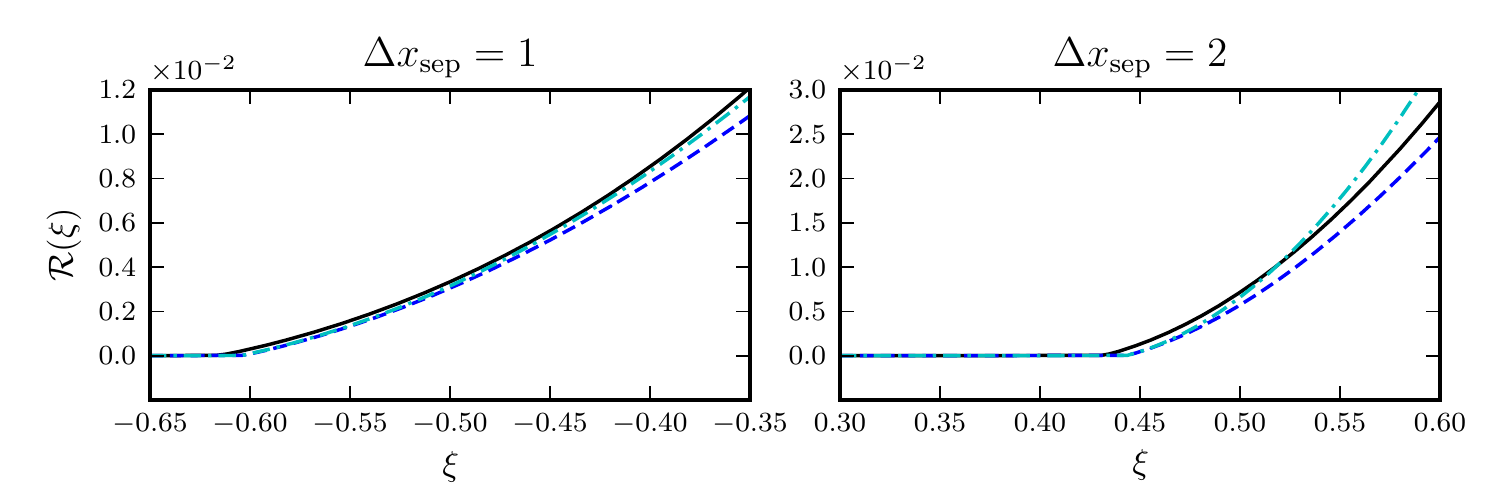} 
   \caption{The comoving curvature perturbation calculated three different ways for the benchmark quartic model. Solid black lines show the perturbation calculated from simulations, as in Sec.~\ref{sec:simulations}. Dashed blue lines show a semi-analytic approximation to the perturbation using Eq.~\ref{eq:any_delta_phi_step} and the simulated bubble profile at the time of the collision. Dash-dotted cyan lines show the fully analytic approximation assuming a step function plus linear profile (Eqs.~\ref{eq:R_analytic_step1} and~\ref{eq:R_analytic_lin1}).
   \label{fig:analytic_perturbations}
   }
\end{figure}

Figure~\ref{fig:analytic_perturbations} shows the final comoving curvature perturbation using several different approximations for the benchmark quartic model. Both the semi-analytic approximation (which uses Eq.~\ref{eq:any_delta_phi_step} and the simulated profile at the time of the collision) and fully analytic approximation (which treats the profile as a step function plus linear piece with $\delta\phi_0 = 1.1\times10^{-3}M_{\rm Pl}$) closely follow the fully numeric calculation near the collision boundary. 
The benchmark collision has nearly unit metric functions and its associated potential is very close to linear in the post-collision region, so we expect the approximations to be valid.
The only noticeable deviation near the collision boundary is a slight offset along the $\xi$-axis between the numeric and analytic solutions. That is, there is a slight offset of the boundary itself. This is due to the analytic approximation that the vacuum energy is equal inside and outside the bubble, whereas the actual collision boundary depends upon the potential parameter $\mu$ (see Eq.~\ref{eq:xi_c}). 
The approximations work best close to the collision boundary where the fields most cleanly superimpose. 

\begin{figure}[t] 
   \centering
   \includegraphics{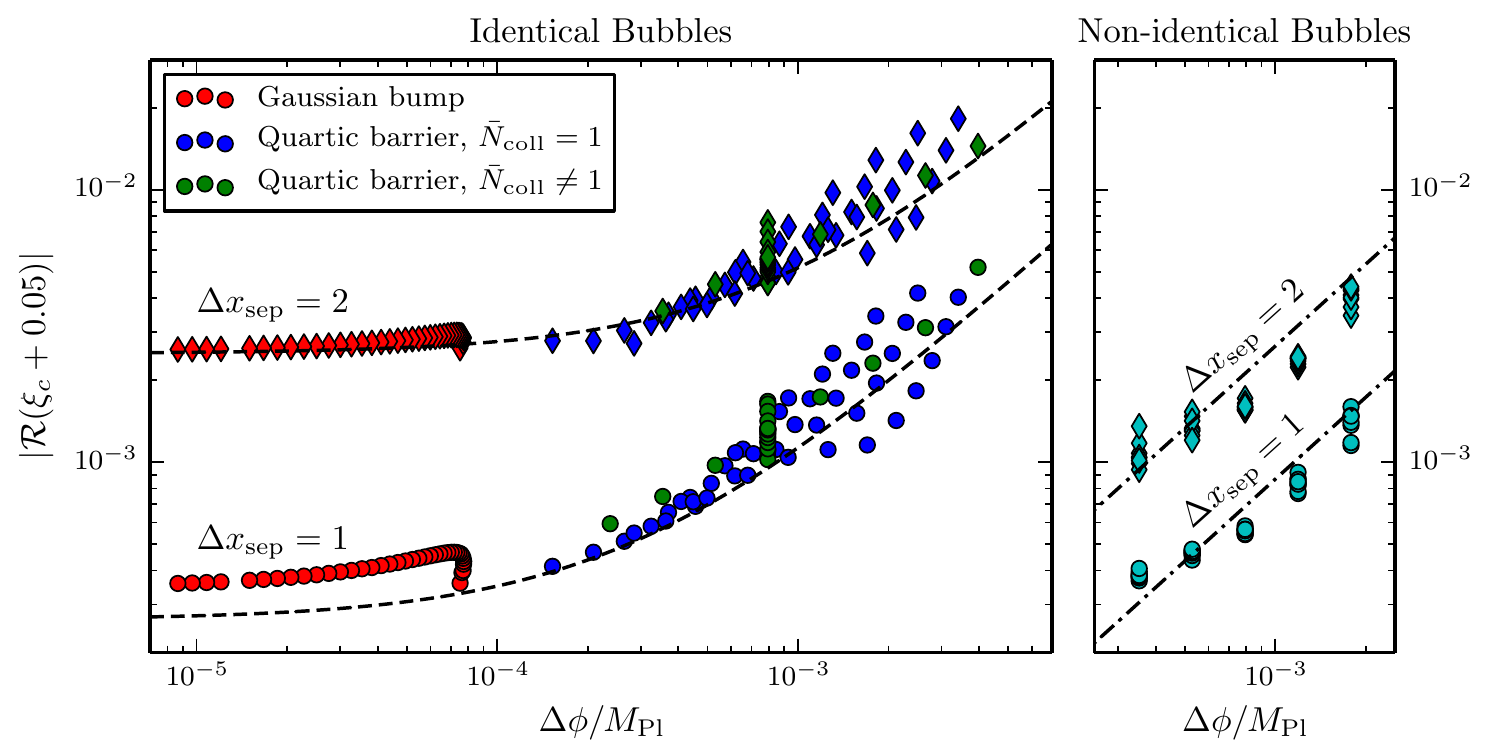} 
   \caption{Scatter plot of the comoving curvature perturbation amplitude for all simulations as a function of the width of the potential barrier. For non-identical bubbles, $\Delta\phi$ denotes the barrier width for the collision bubble. All perturbations are calculated at $\xi-\xi_c = 0.05$, where Eq.~\ref{eq:xi_c} determines $\xi_c$. The two dashed lines show the analytic approximation to the perturbation for identical bubbles ($\mathcal{R} \approx \mathcal{R}_{\rm step} + \mathcal{R}_{\rm lin}$) at $\Delta\xsep=1$ and 2, while the dot-dashed lines show the analytic approximation to the perturbation for non-identical bubbles ($\mathcal{R} \approx \mathcal{R}_{\rm step}$).
   }
   \label{fig:scatterplot}
\end{figure}

In Fig.~\ref{fig:scatterplot} we compare the prediction of Eqs.~\ref{eq:R_analytic_step1} and \ref{eq:R_analytic_lin1} to all of the simulations performed in this paper. 
The analytic approximations (dashed lines) are the sum of $\mathcal{R}_{\rm step}$ and $\mathcal{R}_{\rm lin}$ with $\delta\phi_0$ in Eq.~\ref{eq:R_analytic_step1} given directly by $\Delta\phi$ in the plot. However, as we saw in Sec.~\ref{sec:simulations}, the start of slow-roll inflation can lie significantly beyond end of the barrier ($\delta\phi_0 > \Delta\phi$, where is $\Delta\phi$ is the parameter defining the quartic potential barrier), which accounts for some the scatter in points above the dashed lines. 
The non-identical collision bubbles do not have slow roll interiors, so they are much better approximated by $\mathcal{R}_{\rm step}$ without $\mathcal{R}_{\rm lin}$ (shown as dot-dashed lines).
Taken as a whole, the analytic approximations do an extremely god job of predicting the qualitative and roughly quantitative features of the perturbations.

\subsection{Implications for Observables}

Note that we can rewrite Eq.~\ref{eq:R_analytic_step1} in terms of the first slow-roll parameter $\epsilon$ using $\partial_\tau \phi_{{\rm sr, obs}} = H_I M_{\rm Pl} \sqrt{2 \epsilon}$. We can also relate $\epsilon$ to the tensor-to-scalar ratio in the observation bubble through $r_\obs = 16 \epsilon$, yielding 
\begin{equation}
\label{eq:R_analytic_step2}
\mathcal{R}_{\rm step}(\xi) = \sqrt{\frac{8}{r_\obs}}\frac{\delta\phi_0}{M_{\rm Pl}}
	(1 - \cos \Delta\xsep) (\xi - \xi_c) + \mathcal{O}(\xi-\xi_c)^2.
\end{equation}

In accord with the picture of an increasing center of mass energy with increasing separation, the amplitude of the comoving curvature perturbation increases monotonically with increasing $\Delta \xsep$. The range in $\xi$ on the surface of last scattering accessible to an observer is related to the spatial curvature. For a collision whose boundary is at the position of the observer (creating a half-sky perturbation), we can set $\xi - \xi_c = 2 \sqrt{\Omega_k}$. The maximum amplitude of the observable comoving curvature therefore scales with $\sqrt{\Omega_k}$, or equivalently, exponentially in the number of ``extra" $e$-folds $e^{-N_{\rm extra}}$. A smaller value of the tensor-to-scalar ratio produces a larger amplitude signature. This is because the tensor-to-scalar ratio is related to the excursion in field space during inflation through the Lyth bound, and smaller field excursions are more sensitive to a fluctuation with fixed $\delta \phi$. The width of the potential barriers and epochs of fast-roll enter through $\delta\phi_0$. Holding everything else fixed, wider barriers and longer periods of post-nucleation fast-roll boost the amplitude of the signature.\footnote{
From Eq.~\ref{eq:Nobs}, a large number of observable collisions can result from a large hierarchy between the false vacuum and inflationary energy scales. Simulating potentials with such large hierarchies is computationally unfeasible, and our analytic approximation does not hold. Although we cannot simulate these cases directly, extrapolating our results implies that if the large hierarchy in energy scales occurs over a large enough region in field space, then the amplitude would be large for collisions between identical bubbles.}
With our conventions, in the Sachs Wolfe approximation the temperature anisotropies are related to the comoving curvature perturbation through $\delta T/T = \mathcal{R}/5$.
Identical bubbles have positive $\delta\phi_0$ and $\mathcal{R} > 0$ (which maps to a hot spot in the CMB), whereas non-identical bubbles that tunnel in opposite directions have negative $\delta\phi_0$ and $\mathcal{R} < 0$ (which maps to a cold spot in the CMB). A Lagrangian that predicts both identical and non-identical bubble collisions could therefore produce both hot and cold spots in a single observer's CMB sky.

Let us now return to the interesting conclusion that there is a lower bound on the amplitude of the comoving curvature depending only on the interior cosmologies. For identical bubbles, the amplitude depends only on $\Omega_k$. Noting that the kinematic factor for intermediate values of $\Delta \xsep$ are of order unity, we have the bound $\delta T/ T \gtrsim \Omega_k$. Observational searches for bubble collisions in the CMB yield the rough bound $\delta T/ T \lesssim 10^{-4}$. Therefore, an observation of negative spatial curvature at the level of $10^{-4}$ would rule out all models that predict $\Ncoll > 1$ collisions between identical bubbles! 

Constraints on --- or detection of --- primordial gravitational waves, spatial curvature, and collision signatures provide information on the amplitude of possible collision bubble profiles. This information, together with the constraint that $\Ncoll \sim \mathcal{O}(1)$, provide constraints on or estimates of the parameters in the scalar field Lagrangian.
Small $r$ and large $\Omega_k$ provide the greatest chance of detection of bubble collisions, and therefore the strongest constraints on parameters in the scalar field Lagrangian. 
There will, however, be some residual degeneracy if there are many parameters determining the potential barrier. We will present a full analysis of these constraints using the most recent data from the \emph{Planck} satellite in a future publication.

\section{Discussion and Conclusions}

An explicit link between the scalar potential underlying eternal inflation and possible cosmological observables makes it possible to put quantitative constraints on, or perhaps find evidence for, the eternally inflating Multiverse. In this paper, building on the methodology developed in Ref.~\cite{Wainwright:2013lea}, we have presented a comprehensive numerical and analytic framework for making predictions of the signature of bubble collisions within the context of any single field model of eternal inflation.  This framework is accurate under the following set of approximations:
\begin{itemize}
\item Inflation is driven by a single minimally-coupled scalar field with a potential that does not exhibit a large hierarchy between the energy scale of the false vacuum and the energy scale of slow roll inflation. Resolution is the limiting factor for studying large hierarchies in the numerics. The analytic model treats perturbations to the bubble as test fields in a fixed de Sitter space, and becomes inaccurate in the limit of large hierarchies. Both the analytic and numerical models are easily extendible to multiple scalar fields. 
\item The expected number of observable collisions $\Ncoll$ is of order one. Models where observable collisions are more rare cannot be tested. Models that predict many observable collisions cannot be described accurately using a spacetime with $SO(2,1)$ symmetry, and require a full 3+1 dimensional treatment.
\item The late-time perturbations are in the linear regime over some region of space. In order to extract the comoving curvature perturbation from the synchronous gauge simulations, it is necessary to perform a linear gauge transformation. Because large perturbations are observationally excluded, this is not a severe restriction. 
\item The inflationary model is large-field, with fixed post-nucleation cosmology in the observation bubble. These approximations are for convenience; both can be relaxed.
\end{itemize} 

To accurately reconstruct the cosmological perturbations caused by bubble collisions, we have performed a suite of numerical simulations for two types of models. The Gaussian bump model (Sec.~\ref{sec-gaussbump}) has only two parameters, and allows for collisions between identical bubbles only. The quartic barrier model (Sec.~\ref{sec-quarticbarrier}) has up to six parameters and allows for the collision between identical and non-identical bubbles. In either case we can reduce the parameter count by one by requiring $\Ncoll = 1$.

A striking feature of the simulations is the predictive power of the free-passage approximation~\cite{Giblin:2010bd}: colliding bubbles simply superpose in the aftermath of a collision.  
Under the above assumptions, this result is very general and not tied to further approximations such as the presence of thin bubble walls.
Some general conclusions can be immediately drawn from superposition of bubble profiles:
\begin{itemize}
\item As noted by previous studies~\cite{Gobbetti_Kleban:2012}, in single-field models, collisions between identical bubbles always results in positive $\mathcal{R}$ (hot spots in the CMB), while collisions between non-identical bubbles always results in negative $\mathcal{R}$ (cold spots in the CMB).
\item The greater the amplitude of the bubble profile at the time of the collision, the larger the amplitude of the perturbation; an important finding of this paper is that this amplitude can have important contributions {\em both} from the width of the potential barrier and also from the post-tunneling cosmological evolution.
\item The dependence on the post-tunnelling cosmological evolution sets a minimum perturbation amplitude in the collision between identical bubbles (in the limit of a vanishing barrier width).
\end{itemize}

The numerical simulations yield additional quantitative and qualitative results. For simulations between identical bubbles, the power-law Eq.~\ref{eq:fitting} was consistently a good fit for the comoving curvature perturbation in the vicinity of the collision boundary.  For the Gaussian bump model, the amplitude and power law index did not change appreciably over the entire surveyed range of parameters. This can be accounted for using superposition by noticing that the $\Ncoll = 1$ constraint forces the width of the barrier to be much smaller than the amplitude of the profile from the inner-bubble cosmology, as seen in Fig.~\ref{fig:simulation_lines}. Therefore, a variation in allowed barrier widths lead to a small fractional variation in the signature.

For collisions between identical bubbles in the quartic barrier model, there was a much more significant variation in the fitted amplitude (by about a factor of 3 over the surveyed range) and power law index (which varied between approximately $1 < \kappa < 2.4$). This variation was most strongly correlated with changes in the width of the potential barrier and asymmetries in the barrier which lead to a period of post-tunnelling fast roll before inflation begins. Again, this is predicted by superposition, since both the barrier width and fast roll lead to larger amplitude field profiles. The power law indices and amplitudes obtained for the models studied in this paper are consistent with those obtained in Ref.~\cite{Wainwright:2013lea} for a single model.

Collisions between non-identical bubbles in the quartic barrier model exhibited extra structure due to the excitation of fluctuations on the post-collision domain wall. For this reason, a  power law is not an accurate fit. However, both the sign and overall magnitude of the fluctuations were consistent with the free-passage approximation.

Numerical simulations can be efficiently performed to obtain predictions for specific models or families of models. However, we have also presented a set of analytic approximations for the comoving curvature perturbation produced by bubble collisions. If the inflaton can be treated as a test field, we presented in Eq.~\ref{eq:fourier_transform0} the exact solution for left moving wave packets in de Sitter space with $SO(2,1)$ symmetry. During the early stages of slow-roll inside the observation bubble, using the free passage approximation, we can describe the perturbation caused by the collision as such a wave packet, and solve for the subsequent evolution. An analogous procedure was carried out in Refs.~\cite{Chang_Kleban_Levi:2009,Gobbetti_Kleban:2012} under similar assumptions. 

When they are valid, the analytic expressions Eq.~\ref{eq:R_analytic_step1} and~\ref{eq:R_analytic_lin1} provide an accurate estimate of the signature in the vicinity of the collision boundary, as can be seen from the comparison with numerical simulations in Fig.~\ref{fig:scatterplot}. Not only the overall amplitude of the signature, but also the shape, are in good agreement with numerical results. The power law indices found in fits of numerical simulations lie in the range $\kappa \sim 1-2$. Profiles that are dominated by the potential barrier are described by Eq.~\ref{eq:R_analytic_step1} and have power law indices $\kappa \sim 1$ while profiles that are dominated by the post-tunnelling cosmological evolution are described by Eq.~\ref{eq:R_analytic_lin1} and have power law indices $\kappa \sim 2$. A general profile will lie somewhere between these values. Finally, we confirmed the numerical result that an increasing de Sitter invariant distance between the bubble centers leads to a larger amplitude perturbation.

By mapping out how changes in the Lagrangian affect the collision phenomenology, these analyses have allowed us to identify the parameters that are most phenomenologically relevant. Two of these parameters, the observed energy density in curvature ($\Omega_k$) and the ratio of the tensor power to the scalar power of primordial fluctuations ($r$), depend only on the observation bubble's cosmology. The other relevant parameters in the Lagrangian are those  determining the distance in field space between the false vacuum and the location where post-nucleation slow-roll evolution begins, such as the width of the barrier surrounding the false vacuum. Finally, the de Sitter invariant distance between the bubble centers (a stochastic variable) determines the overall amplitude. An interesting relation between these parameters, illustrative of constraints that might be imposed on theories of eternal inflation, is 
that the minimum amplitude for perturbations between identical bubbles depends only upon $\Omega_k$: $\delta T / T > \Omega_k (1- \cos\Delta x_{\rm sep})^2 / 10$. 
An observation of spatial curvature would therefore be extremely constraining. 

Determining the full constraining power of bubble collisions on the scalar field Lagrangian underlying eternal inflation will be the topic of future work.

\acknowledgments 

This work was partially supported by a New Frontiers in Astronomy and Cosmology grant \#37426. 
Research at Perimeter Institute is supported by the Government of Canada through Industry Canada and by the Province of Ontario through the Ministry of Research and Innovation. 
MCJ is supported by the National Science and Engineering Research Council through a Discovery grant. 
AA was supported in part by a time release grant from the Foundational Questions Institute (FQXi), of which he is Associate Director. 
HVP is supported by STFC and the European Research Council under the European Community's Seventh Framework Programme (FP7/2007-2013) / ERC grant agreement no 306478-CosmicDawn.

\appendix
\section{Fourier Transforms}
\label{app:fourier_transforms}

The step from Eq.~\ref{eq:fourier_transform0} to Eq.~\ref{eq:fourier_transform} contains a non-trivial amount of algebra. Intuitively, Eq.~\ref{eq:fourier_transform} represents a sum over independently evolving Fourier modes with initial conditions given at time $N_0$. We now show this more rigorously.

Let
\begin{equation}
\Phi(N,k) = \frac{1}{2\pi}\int \phi(\tilde{x},N) e^{ik\tilde{x}}d\tilde{x}
\end{equation}
be the time-dependent Fourier transform of the field $\phi$ (the limits of all integrals in this section are implicitly set to $(-\infty, \infty)$). Then the field $\Phi$ satisfies Eq.~\ref{eq:homo_eom}, with general solutions given by 
\begin{equation}
\Phi(N,k) = F_1(k) \phi_A(N, N_0, k) + F_2(k) \phi_B(N, N_0, k),
\end{equation}
where by construction $\phi_A(N_0) = \dot{\phi}_B(N_0) = 1$ and $\dot\phi_A(N_0) = \phi_B(N_0) = 0$. Therefore, 
\begin{equation}
F_1(k) = \Phi(N_0, k)\text{ and }F_2(k) = \dot\Phi(N_0, k).
\end{equation}

Near $N=N_0$, we wish to find solutions that are traveling with a uniform velocity $c_0$. That is, $\phi(N,x) \approx f(x-c_0 N)$. This implies that $\dot\phi(N_0,x) = -c_0 \partial_x \phi(N_0, x)$ and $\dot\Phi(N_0, k) = ikc_0 \Phi(N_0, k)$, so 
\begin{equation}F_2(k) = ikc_0 F_1(k).\end{equation}
Taking the inverse Fourier transform yields
\begin{align}
\phi(N,x) &= \int \Phi(N, k) e^{-ikx}\,dk \\
	&= \int \Phi(N_0, k) \left[\phi_A + ikc_0 \phi_B\right] e^{-ikx}\,dk \\
	&= \frac{1}{2\pi} \int\int \phi(\tilde{x}, N_0) \left[\phi_A + ikc_0 \phi_B\right] e^{-ik(x-\tilde{x})}\,dk\,d\tilde{x}.
\end{align}
Since both $\phi_A$ and $\phi_B$ are even in $k$, the odd (imaginary) terms cancel over even limits of integration, leaving us with
\begin{equation}
\phi(N,x) = \frac{1}{2\pi} \int\int \phi(\tilde{x}, N_0) \left[\cos(k(x-\tilde{x})) \phi_A + c_0 k \sin(k(x-\tilde{x})) \phi_B\right] \,dk\,d\tilde{x},
\end{equation}
which, after a change in the limits of integration, is the same as Eq.~\ref{eq:fourier_transform}.

\section{Transforming a Step Function}
\label{app:step_transform}
To integrate Eq.~\ref{eq:fourier_transform} with $\phi(N_0, x)$ equal to a step function, we must first find the Fourier transform of a step function:
\begin{equation}
\frac{1}{\pi}\int_{-\infty}^\infty  \Theta (\tilde{x}) e^{ik(x-\tilde{x})}\, d\tilde{x} = \left[ \delta(k) - \frac{i}{\pi k}\right] e^{ikx}.
\end{equation}
Therefore,
\begin{multline}
\frac{1}{\pi} \int_{-\infty}^\infty \Theta (\tilde{x}) [ \cos(k(x-\tilde{x}))\phi_A + B \sin(k(x-\tilde{x}))\phi_B]  \, d\tilde{x}\\
	= \left[\delta(k) \cos kx + \frac{\sin kx}{\pi k}\right]\phi_A
	+ \left[\delta(k) \sin kx - \frac{\cos kx}{\pi k}\right]\phi_B.
\end{multline}
By plugging in the asymptotic expressions for $\phi_A$ and $\phi_B$ (Eqs.~\ref{eq:phiA_asym} and~\ref{eq:phiB_asym}) and integrating over $k$, we find a solution in terms of the integrals
\begin{align}
\int_0^\infty \frac{\cos (k\eta) \sin (kx)}{k (1-k^2)} dk &= \frac{\pi}{2} \sin\eta \sin x \\
\int_0^\infty \frac{\sin (k\eta) \cos (kx)}{k (1-k^2)} dk &= \frac{\pi}{2} (1 - \cos\eta \cos x) \\
\int_0^\infty \frac{\sin (k\eta) \sin (kx)}{1-k^2} dk &= -\frac{\pi}{2} \cos\eta \sin x \\
\int_0^\infty \frac{\cos (k\eta) \cos (kx)}{1-k^2} dk &= +\frac{\pi}{2} \sin\eta \cos x.
\end{align}
Each can be solved with contour integration by assuming that $0 < |x| < \eta$. To get the solutions for $|x| > \eta > 0$, one can simply switch the dummy variables $x$ and $\eta$ in the integration. Using these, it is straightforward to derive Eq.~\ref{eq:step_solution}.

\bibliographystyle{JHEP}
\bibliography{bubble_pheno}

\end{document}